\documentclass[11pt]{article}

\input{preamble}
\usepackage{comment}
\usepackage{ifthen}

\ifthenelse{\isundefined{\workingversion}}{%
\excludecomment{trash}
\excludecomment{workingversion}
\renewcommand{\marginpar}[1]{}
}%
{
\inputcomment{trash}
\inputcomment{workingversion}
}

\begin{document}

\title{Completeness for Flat Modal Fixpoint Logics\thanks{
    Research supported by the Van Gogh research project \emph{Modal
      Fixpoint Logics}. 
  }
}

\newcommand{\email}[1]{\emph{Email:}~\texttt{#1}}

\newcommand{\luigisaddress}{Laboratoire d'Informatique Fondamentale de
  Marseille, Universit\'e de Provence, 39 rue F. Joliot Curie, 13453
  Marseille Cedex 13, France.
  \email{luigi.santocanale@lif.univ-mrs.fr}} 

\newcommand{\ydesaddress}{%
  Institute for Logic, Language and Computation, Universiteit van
  Amsterdam, 
  Plantage
  Muidergracht 24, 1018 TV Amsterdam, Netherlands.
  \email{Y.Venema@uva.nl} 
}

\author{Luigi Santocanale\thanks{\luigisaddress}
  \and Yde Venema\thanks{\ydesaddress}}

\date{\today}

\maketitle
 
\begin{abstract}
  This paper exhibits a general and uniform method to prove completeness
  for
  certain modal fixpoint logics.  Given a set $\Ga$ of modal
  formulas of the form $\ga(x,p_1,\ldots,p_n)$, where $x$ occurs only
  positively in $\ga$, the language $\langfx(\Ga)$ is obtained
  by adding to the language of polymodal logic a connective
  $\ff_{\ga}$ for each $\ga \in \Ga$. The term
  $\ff_{\ga}(\phi_{1},\ldots,\phi_{n})$ is meant to be interpreted as the least
  fixed point of the functional interpretation of the term
  $\ga(x,\phi_{1},\ldots,\phi_{n})$. We consider the following problem: given
  $\Ga$, construct an axiom system which is sound and complete with
  respect to the concrete interpretation of the language
  $\langfx(\Ga)$ on Kripke frames.  We prove two results that solve
  this problem.  

  First, let $\Klogfx(\Ga)$ be the logic obtained
  from the basic polymodal $\Klog$ by adding a Kozen-Park style
  fixpoint axiom and a least fixpoint  rule, for each fixpoint
  connective $\ff_{\ga}$. Provided that each indexing formula $\ga$
  satisfies the syntactic criterion of being untied in $x$, we prove
  this axiom system to be complete.

  Second, addressing the general case, we prove the
  soundness and completeness of an extension $\Klogfxp(\Ga)$ of
  $\Klogfx(\Ga)$. This extension is obtained via an effective
  procedure that, given an indexing formula $\ga$ as input, returns a
  finite set of axioms and  derivation rules for $\ff_{\ga}$, of size
  bounded by the length of $\ga$. Thus the axiom system
  $\Klogfxp(\Ga)$ is finite whenever $\Ga$ is finite. 
  \vskip 2pt  
\noindent
{\bf Keywords.} fixpoint logic, modal logic, axiomatization,
completeness,  least fixpoint, modal algebra, representation theorem
\end{abstract} 

\section{Introduction}
\label{s:1}

\newcommand{\vphi}{\phi_{1},\ldots ,\phi_{n}}
\newcommand{\vp}{p_{1},\ldots ,p_{n}}

Suppose that we extend the language of basic (poly-)modal logic with a
set $\set{ \ff_{\ga} \mid \ga \in \Ga }$ of so-called \emph{fixpoint
  connectives}, which are defined as follows.  Each connective
$\ff_{\ga}$ is indexed by a modal formula $\ga(x,\vp)$ in which
$x$ occurs only positively.
The intended meaning of the formula $\ff_{\ga}(\vphi)$ in a
labelled transition system (Kripke model) is the least fixpoint of the
formula $\ga(x,\vphi)$,
\[
\ff_{\ga}(\vphi) \equiv \mu x.\ga(x,\vphi).
\]
Many logics of interest in computer science are of this kind: Such
fixpoint connectives can be found for instance in $\PDL$,
propositional dynamic logic~\cite{kozen:PDL}, in $\CTL$, computation
tree logic~\cite{emerson}, in $\LTL$, linear temporal logic, and in
multi-agent versions of epistemic 
logic~\cite{faginhalpernetal}.  
More concretely, the Kleene iteration diamond $\exop{a^{*}}$ of $\PDL$ can
be presented (in the case of an atomic program $a$) as the
connective $\ff_{\delta}$, where $\delta(x,p)$ is the formula $p \lor
\exop{a}x$: the formula $\exop{a^{*}}\phi$ can be interpreted as the
parameterized least fixpoint $\mu_{x}.\delta(x,\phi)$.  As two more
examples, 
let $\theta(x,p,q) := p \vee (q \land \pos x)$, and $\eta(x,p,q) := p
\vee (q \land \nec x)$, then $\CTL$ adds new connectives
$\ff_{\theta}(p,q),\ff_{\eta}(p,q)$ --- or $E(p\,Uq), A(p\,Uq)$ in the
standard notation --- to the basic modal language.

Generalizing these examples we arrive at the notion of a flat modal
fixpoint logic.  Let $\langfx(\Ga)$ denote the language we obtain if
we extend the syntax of (poly-)modal logic with a connective
$\ff_{\ga}$ for every $\ga \in \Ga$.  Clearly, every fixpoint
connective of this kind can be seen as a macro over the language of
the modal $\mu$-calculus.  Because the associated formula $\ga$ of a
fixpoint connective is itself a basic modal formula (which explains
our name \emph{flat}), it is easy to see that every flat modal
fixpoint language is contained in the alternation-free fragment of the
modal $\mu$-calculus~\cite{kozen}.  Because of their transparency and
simpler semantics, flat modal fixpoint logics such as $\CTL$ and
$\LTL$ are often preferred by end users.  In fact, most verification
tools implement some flat fixpoint logic rather than the full
$\mu$-calculus, regardless of considerations based on the expressive
power of these logics.
\medskip

Despite their wide-spread applications and mathematical interest, up to now
general investigations of modal fixpoint logics have been few and far between.
In this paper we address the natural problem of \emph{axiomatizing} flat modal
fixpoint logics.
Here the flat modal fixpoint \emph{logic} induced by $\Gamma$ is the set of
$\langfx(\Ga)$-validities, that is, the collection of formulas in the language
$\langfx(\Ga)$ that are true at every state of every Kripke model.

In general, the problem of axiomatizing fixpoints arising in computer
science is recognized to be a nontrivial one.  As an example we
mention the longstanding problem of axiomatizing regular
expressions~\cite{conway,boffa,krob,kozen:regexpr}, whereas the
monograph%
~\cite{bloomesik} is a good general survey on fixpoint theory.  More
specifically, in the literature on modal logic one may find
completeness results for a large number of individual systems.  We
mention the work of Segerberg~\cite{sege:comp82} and of Kozen \&
Parikh~\cite{koze:elem81} on PDL, the axiomatization of Emerson \&
Halpern~\cite{emer:deci85} of CTL, and many results on epistemic logic
with the common knowledge operator or similar
modalities~\cite{faginhalpernetal,meye:epis95}.  In the
paper~\cite{kozen} that introduced the modal $\mu$-calculus, Kozen
proposed an axiomatization which he proved to be complete for a
fragment of the language; the completeness problem of this
axiomatization for the full language was solved positively by
Walukiewicz~\cite{walukiewicz}.  But to our knowledge, no general
results or uniform proof methods have been established in the theory
of modal fixpoint logics.  For instance, the classical filtration
methods from modal logic work for relatively simple logics such as
$\PDL$~\cite{kozen:PDL}, but they already fail if this logic is
extended with the loop operator~\cite{kozen}.  A first step towards a
general understanding of flat fixpoint logics  is the work
\cite{lang:focu01}, where a game-based approach is developed to deal
with axiomatization and satisfiability issues for $\LTL$ and $\CTL$.

In this paper we contribute to the general theory of flat modal fixpoint
logics by providing completeness results that are uniform in the parameter
$\Ga$, and modular in the sense that the axiomatizations take care of each
fixpoint connective separately. 
Our research is driven by the wish to understand the combinatorics of
fixpoint logics in their wider mathematical setting. 
As such it continues earlier investigations by the first author into the
algebraic and order-theoretic aspects of fixpoint calculi%
~\cite{santocanale:LICS05,sant:comp08}, and work by the second author
on coalgebraic (fixpoint)
logics~\cite{vene:auto06,kupk:comp08,kupk:coal08}.

\medskip

Usually, the difficulty in finding a complete axiomatization problem
for a fixpoint logic does not stem from the absence of a natural
candidate.  In our case, mimicking Kozen's axiomatization of the modal
$\mu$-calculus, an intuitive axiomatization for the
$\langfx(\Ga)$-validities would be to add, to some standard
axiomatization $\Klog$ for (poly-)modal logic, an axiom for each
connective $\ff_{\ga}$ stating that $\ff_{\ga}(\vp)$ is a prefixpoint
of the formula $\ga(x,\vp)$, and a derivation rule which embodies the
fact that $\ff_{\ga}(\vp)$ is the smallest such.

\begin{definition}
\label{d:ax1}
The axiom system $\Klogfx(\Ga)$ is obtained by adding to $\Klog$ the axiom
\begin{equation}
\label{ax:sharpprefix}%
\tag{$\ff_{\gamma}$-prefix}%
\ga(\ff_{\gamma}(\vp),\vp) \to \ff_{\gamma}(\vp), %
\end{equation}
and the derivation rule\footnote{\label{fn:1}%
   This rule is to be interpreted as stating that if some
   substitution instance $\ga(\psi,\vphi) \to \psi$ of the premiss
   is derivable in the system, then so is the corresponding substitution
   $\ff_{\ga}(\vphi) \to \psi$ of the conclusion.
   Algebraically, it corresponds to the quasi-equation 
   $\ga(y,\vp) \leq y \;\to\; \ff_{\ga}(\vp) \leq y$
   (or to the Horn formula obtained from this quasi-equation by universally
   quantifying over the variables $y$ and $\vp$).
   }
\begin{equation}
  \label{ax:sharpleast}%
  \tag{$\ff_{\gamma}$-least}%
\frac{ \ga(y,\vp) \to y}{ \ff_{\ga}(\vp) \to y}
\end{equation}
for each $\ga \in \Ga$.
\end{definition}

In fact, the first of our two main results, Theorem~\ref{t:scs}, states that
for many  choices of $\Gamma$, $\Klogfx(\Ga)$ is indeed a complete
axiomatization.
More precisely, we identify a class of formulas that we call \emph{untied
in $x$} --- these formulas are related to the aconjunctive~\cite{kozen}
and disjunctive~\cite{walukiewicz} formulas from the modal $\mu$-calculus. 
In this paper we shall prove that
\begin{center}
if every $\ga$ in $\Ga$ is untied in $x$, then $\Klogfx(\Ga)$ is a complete 
axiomatization.
\end{center}
This result takes care of for instance the completeness of $\CTL$.

However, the road to a general completeness result for the
system $\Klogfx(\Ga)$ is obstructed by a familiar problem, related to the role
of conjunctions in the theory of fixpoint logics.
Our solution to this problem comprises a \emph{modification} of the intuitive
Kozen-style axiomatization, inspired by a construction of Arnold \&
Niwi\'nski~\cite{arnoldniwinski}.
Roughly speaking, this so-called Subset Construction is a procedure that
simulates a suitable system of equations $T$ by a system of equations
$T^{+}_{\ga}$ that we will call \emph{simple} since it severely restricts
occurrences of the conjunction symbol.
It is shown in~\cite[\S 9.5]{arnoldniwinski} that on \emph{complete} lattices,
the least solutions of $T$ and $T^{+}$ may be constructed from one another.
The key idea of our axiomatization is first to represent $\ga$ by an equivalent
system of equations $T_{\ga}$, and then to \emph{force} the simulating system
$T^{+}_{\ga}$ to have a least solution, constructible from $\ff_{\ga}$, on
the algebraic models for the logic.

More concretely, we present a simple algorithm that produces, when given as 
input a modal formula $\ga(x)$ that is positive in $x$, a finite set of 
axioms and rules, of bounded size.
Adding these axioms and rules to the basic modal logic $\Klog$, we obtain
an axiom system $\Klogfxp(\Ga)$, which is finite if $\langfx(\Ga)$ has
finitely many fixpoint connectives.
Our second main result, Theorem~\ref{t:sc}, states that, for \emph{any} flat
fixpoint language, 
\begin{center}
$\Klogfxp(\Ga)$ is a complete axiomatization for the validities in 
$\langfx(\Ga)$.
\end{center}

Let us briefly describe the strategy for obtaining the completeness
theorem. 
We work in an algebraic setting for modal logic.
Following a well known approach of algebraic logic, we treat formulas as
terms over a signature whose function symbols are the logical connectives.
Then, axioms correspond to equations and
derivation rules to quasi-equations. The algebraic counterpart of the
completeness theorem states that the equational theory of the ``concrete"
algebraic models that arise as complex algebras based on Kripke frames, 
is the same of the equational theory of the algebraic models of our 
axiomatization. 
To obtain such an algebraic completeness theorem, we study the
Lindenbaum-Tarski algebras of our logic. 
Two properties of these structures turn out to be crucial: 
First, we prove that every Lindenbaum-Tarski algebra is \emph{residuated},
or equivalently, that every diamond of the algebra has a right adjoint.
And second, we show that the Lindenbaum-Tarski algebras are 
\emph{constructive}: every fixpoint operation can be approximated as the
join of its finite approximations.
Then, we prove an algebraic representation theorem, Theorem~\ref{t:4:1},
stating that every countable algebra with these two properties can be
represented as a Kripke algebra, that is, as a subalgebra of the complex
algebra of a Kripke frame.
Putting these observations together, we obtain that the countable
Lindenbaum-Tarski algebras have the same equational theory as the Kripke
algebras, and this suffices to prove the algebraic version of the
completeness theorem.

In order to prove these remarkable properties of the Lindenbaum-Tarski
algebras, we switch to a \emph{coalgebraic} reformulation of modal
logic, based on the coalgebraic or cover modality $\nb$.  This
connective $\nb$ takes a finite \emph{set} $\alpha$ of formulas and
returns a single formula $\nb\alpha$, which can be seen as the
following abbreviation:
\[
\nabla\alpha = \Box(\bigvee \alpha)\wedge \bigwedge
\Diamond \alpha,
\]
where $\Diamond\alpha$ denotes the set $\set{ \Diamond a \mid a \in \alpha }$.
The pattern of the definition of $\nb$ has surfaced in the literature on modal
logic, in particular, as Fine's normal forms~\cite{fine:norm75}.
The first explicit occurrences of this modality as a primitive connective,
however, appeared not earlier than the 1990s, in the work of Barwise \&
Moss~\cite{barw:vici96} and of Janin \& Walukiewicz~\cite{jani:auto95}.
We call this connective ``coalgebraic'', because of Moss' 
observation~\cite{moss:coal99}, that its semantics allows a natural formulation
in the framework of Universal Coalgebra, a recently emerging general mathematical
theory of state-based evolving systems~\cite{rutt:univ00}.
Moss' insight paved the way for the transfer of many concepts, results and
methods from modal logic to a far wider setting.
As we will see, the main technical advantage of reconstructing modal logic on
the basis of the cover modality is that this allows one to, if not completely
eliminate conjunctions from the language, then at least \emph{tame} them, so
that they become completely harmless.
This reduction principle, which lies at the basis of many constructions in
the theory of the modal $\mu$-calculus~\cite{jani:auto95}, has recently been
investigated more deeply~\cite{palm:nabl07,bilk:proo08}, and generalized to
a coalgebraic level of abstraction~\cite{kupk:clos05,kupk:comp08}.

We now briefly discuss how the present work contributes to the existing theory
of fixpoint logics.  
Perhaps the first observation should be that our completeness results does
not follow from Walukiewicz' completeness result for the modal 
$\mu$-calculus~\cite{walukiewicz}: each \emph{language} $\langfx(\Ga)$
may be a fragment of the full modal $\mu$-calculus, but this does not imply
that Kozen's axiomatization of the modal $\mu$-calculus is a 
\emph{conservative extension} of its restriction to such a language.
In this respect, our results should be interpreted by saying that we add to
Walukiewicz' theorem the observation that, modulo a better choice of axioms,
proofs of validities in any given flat fragments of the modal $\mu$-calculus
can be carried out \emph{inside} this fragment.

And second, while our methodology is based on earlier work~\cite{sant:comp08}
by the first author, which deals with the alternation-free fragment of the
$\mu$-calculus, we extend these results in a number of significant ways.
In particular, the idea to use the subset
construction of Arnold \& Niwi\'nski to \emph{define an axiom system}
for flat modal fixpoint logics, is novel.  Furthermore, the
representation theorem presented in Section \ref{sec:embedding}
strengthens the main result of \cite{sant:comp08} (which applies to
complete algebras only), to a completeness result for \emph{Kripke
  frames}.
With respect to \cite{sant:comp08}, we also emphasize here the role of the
coalgebraic cover modality $\nb$ in the common strategy for obtaining
completeness.
It is not only that some obscure results of~\cite{sant:comp08} get a specific
significance when understood from the coalgebraic perspective, but we also
prove some new results on the cover modality $\nb$ itself, which may be of
independent interest.
And lastly, we can place an observation similar to the one we made with 
respect to Walukiewicz' result for the full modal $\mu$-calculus: the results 
in~\cite{sant:comp08} do not necessarily carry over to arbitrary
fragments that are flat fixpoint logics.
In fact, we were surprised to observe that it turns out to be possible to find
a finitary complete axiomatization of the fixpoint connective $\ff_{\gamma}$
without explicitly introducing in the signature the least fixpoint of some
other formula $\delta$. This fact contrasts with the method proposed in
\cite{santocanale:eqfixedpoints} to equationally axiomatize the prefixpoints.

Finally, our proof method and, consequently, all of our results apply to
the framework of \emph{polymodal} logic, and we have formulated our main
results accordingly.
However, since much of the material presented here requires some rather
involved notation, we will frequently choose to work in the setting of
\emph{monomodal} logic, in order to keep the text as readable as possible.
In those cases where the transition to the polymodal setting is not routine,
we always provide explicit details of this transition.

\paragraph{Overview of the paper.}
In Section \ref{s:preliminaries}
we first define flat modal fixpoint logics and then introduce our main 
tools: the coalgebraic cover modality $\nb$, the algebraic approach to modal
(fixpoint) logic, the order theoretic notion of a finitary $\O$-adjoint, and
the concept of a system of equations.
Section~\ref{s:ax} is devoted to the axiomatization $\Klogfxp(\Ga)$ which we
present as an algorithm producing the axiomatization given as input a set
$\Gamma$ of modal formulas.
In Section~\ref{s:soe} we give the proof of some algebraic results that
relate fixpoints of different functions and that are at the core of the
axiomatizations $\Klogfx(\Ga)$ and $\Klogfxp(\Ga)$.
With these results at hand, in Section~\ref{s:sc} we formulate our two
soundness and completeness results, and we sketch an overview of our
algebraic proof method, introducing the Lindenbaum-Tarski algebras $\Li$.
In Section~\ref{s:constructive}, we show that these Lindenbaum-Tarski
algebras $\Li$ have a number of properties that make them resemble the 
power set algebra of a Kripke frame:  we prove $\Li$ successively to be
rigid, residuated, and constructive.
Finally, in Section~\ref{s:embedding}, we prove the above-mentioned 
representation theorem stating that every countable, residuated and 
constructive algebraic model of our language can be represented as a
subalgebra of a powerset algebra of some Kripke frame.


\section{Preliminaries}
\label{s:2}
\label{s:preliminaries}

In this section we present some material that we consider background
knowledge in the remainder of the paper.
We first give a formal definition of the syntax and semantics of flat
modal fixpoint logics.  We then discuss the reformulation of modal
logic in terms of the cover modalities $\nab[i]$.  Finally, we
introduce modal $\ff$-algebras as the key structures of the algebraic
setting in which we shall prove our completeness result.  For
background in the algebraic perspective on modal logic,
see~\cite{blac:moda01,vene:alge06}.

\subsection*{Flat modal fixpoint logic}
The flat modal fixpoint logic of language $\langfx(\Ga)$ will be an
extension of polymodal logic.
Therefore we shall use $I$ to denote the finite set of atomic
actions indexing the modalities of polymodal logic.
Next -- and throughout this paper -- we fix a set $\Ga$ of polymodal
formulas $\ga(x,\vec{p})$ where the variable $x$ occurs only
positively in $\gamma$ and $\vec{p} = (p_{1},\ldots,p_{n})$ is the ordered
list of free
variables in $\gamma$ that are distinct from $x$.  As usual $x$ occurs
only positively in $\gamma$ if each occurrence of $x$ appears under an
even number of negations.  Alternatively, we may decide to present the
syntax of polymodal logic so that negation applies to propositional
variables only, in which case $x$ occurs positively if it occurs under
no negation.
The vector $\vec{p}$ might be different for each $\ga$, but we decided
not to make this explicit in the syntax, in order not to clutter up
notation.

First we give a formal definition of the language of flat modal
fixpoint logics.  Basically we add a new logical connective
$\ff_{\ga}$ to the language, for each $\ga \in \Ga$.
\begin{definition}
  The set $\langfx(\Ga)$ of flat modal fixpoint formulas
  associated with $\Ga$ is defined by the following grammar:
  \[
  \phi ::= p \mid \neg\phi \mid \phi_{1}\land\phi_{2} \mid \mop{i}\phi
  \mid \ff_{\ga}(\vec{\phi})\,,
  \]
  where $p \in P$ is a propositional variable, $i$ and $\ga$ range
  over $I$ and $\Ga$, respectively, and $\vec{\phi}$ is a vector of
  previously generated formulas indexed by the vector $\vec{p}$.
\end{definition}

We move on to the intended semantics of this language.  A
\emph{labeled transition system} of type $I$, or equivalently a
\emph{Kripke model}, is a structure $\bbS = \struc{S, \set{R_{i} \mid
    i \in I}}$, where $S$ is a set of states and, for each $i \in I$,
$R_{i} \subseteq S \times S$ is a transition relation.  

\begin{definition}
Given a Kripke model $\bbS$ and a valuation $\vec{v} : P \rTo \powerset(S)$
of propositional variables as subsets of states, we inductively define
the semantics of flat modal fixpoint formulas as follows:
  \begin{align}
    \nonumber \interpret[\vec{v}]{p} & = \vec{v}(p)\,,\\
    \nonumber \interpret[\vec{v}]{\neg \phi} & =
    \mycomplement{\interpret[\vec{v}]{\phi} }\,,\\
    \nonumber \interpret[\vec{v}]{\phi_{1} \land \phi_{2}} & =
    \interpret[\vec{v}]{\phi_{1}} \cap \interpret[\vec{v}]{\phi_{2}}
    \,,\\
    \nonumber \interpret[\vec{v}]{\pos[i]\phi} & = \set{x \in S \mid
      \exists y \in S \tst xR_{i}y \tand y \in
      \interpret[\vec{v}]{\phi}} \,.\\
    \intertext{%
      In order to define
      $\interpret[\vec{v}]{\ff_{\gamma}(\vec{\phi})}$, let $x$ be a
      variable which is not free in $\vec{\phi}$ and, for $Y \subseteq
      S$, let $(\vec{v},x \rightarrow Y)$ be the valuation sending $x$
      to $Y$ and every other variable $y$ to $\vec{v}(y)$. We let }
    \label{eq:intsharp}
    \interpret[\vec{v}]{\ff_{\gamma}(\vec{\phi})}
    & = 
    \bigcap \set{Y \mid \interpret[(\vec{v},x \rightarrow
      Y)]{\gamma(x,\vec{\phi})} \subseteq Y}\,.
  \end{align}
\end{definition}

Observe that, by the Knaster-Tarski theorem \cite{tarski}, 
\eqref{eq:intsharp} just says that the interpretation of
$\ff_{\gamma}(\vec{\phi})$ is the least fixpoint of the order
preserving function sending $Y$ to $\interpret[(\vec{v},x \rightarrow
Y)]{\gamma(x,\vec{\phi})}$.

\subsection*{The cover modality}

We will frequently work in a reformulation of the modal language based
on the \emph{cover modality} $\nb$.  
This connective, taking a \emph{set} of formulas as their argument, can be
defined in terms of the box and diamond operators:
\[
\nb\Phi := \Box\bigvee\Phi \land \bigwedge\dia\Phi\,,
\]
where $\dia\Phi$ denotes the set $\{ \dia\phi \mid \phi\in\Phi \}$.
Conversely, the standard diamond and box modalities can be defined in terms
of the cover modalities:
\begin{align}
  \label{eq:nb0}
  \dia\phi &\equiv \nb\{\phi,\top\}\,,&
  \Box\phi &\equiv \nb\nada \lor \nb\{\phi\}\,.
\end{align}
It follows from these observations that we may equivalently base our
modal language on $\nb$ as a primitive symbol.

What makes the cover modality $\nb$ so useful is that it satisfies two
distributive laws:

\begin{equation}
\label{eq:nb2}
\nb(\Phi\cup \{ \bv\Psi \}) =
\bigvee_{\emptyset \subset \Psi'\subseteq \Psi} \nb (\Phi\cup\Psi')\,,
\end{equation}
and
\begin{equation}
\label{eq:nb1}
\nb\Phi \land \nb\Psi \equiv 
\bigvee_{Z\in\Phi\bowtie\Psi} \nb \set{ \phi\land\psi \mid (\phi,\psi)\in Z }\,,
\end{equation}
where $\Phi\bowtie\Psi$ denotes the set of relations $R
\sse\Phi\times\Psi$ that are \emph{full} in the sense that for all
$\phi\in\Phi$ there is a $\psi \in\Psi$ with $(\phi,\psi)\in R$, and
vice versa.  The principle (\ref{eq:nb2}) clearly shows how the cover
modality distributes over disjunctions, but we also call
(\ref{eq:nb1}) a distributive law since it shows how conjunctions
distribute over $\nb$.

\begin{remark}
  For more information on these distributive laws, the reader is
  referred to~\cite{palm:nabl07,bilk:proo08}, or
  to~\cite{kupk:comp08}, where these principles are shown to hold in a
  very general coalgebraic context.  Although to our knowledge it has
  never been made explicit in the literature on automata theory,
  equation~\eqref{eq:nb1} is in fact the key principle allowing the
  simulation of alternating automata by non-deterministic ones within
  the setting of $\mu$-automata~\cite{jani:auto95}.  We refer
  to~\cite{jani:auto97} for an algebraic, or
  to~\cite{kupk:clos05,kupk:coal08} for a coalgebraic explanation of
  this.
\end{remark}

As a straightforward application of these distributive laws (together with
the standard distribution principles of conjunctions and disjunctions), every
modal formula can be brought into a normal form, either by pushing conjunctions
down to the leaves of the formula construction tree, or by pushing disjunctions 
up to the root, or by doing both.
In order to make this observation more precise, we need some definitions,
where we now switch to the polymodal setting in which we have a cover 
modality $\nab[i]$ for each atomic action $i$.

\begin{definition}
Let $X$ be sets of propositional variables.  Then we define the following
sets of formulas:
\begin{enumerate}
\item 
$\Lit(X)$ is the set $\set{ x, \neg x \mid x \in X }$ of \emph{literals} over $X$,
\item 
$\MT(X)$ is the set of \emph{$\nb$-formulas} over $X$ given by the following grammar:
\[
\phi ::= x \mid \neg x \mid  \bot \mid \phi\lor\phi \mid
    \top \mid \phi\land\phi
    \mid \nab[i]\Phi
\]
where $x\in X$, $i \in I$, and $\Phi \sse \MT(X)$.    
\item
$\MTNF(X)$ is the set of \emph{disjunctive} formulas given by the 
 following grammar:
\[
\phi ::=  \bot \mid \phi\lor\phi \mid
    \mbox{$\bigwedge\Lambda$}\land\bigwedge_{j \in J}\nab[j]\Phi_{j},
\]
where $\Lambda \sse \Lit(X)$, $J \subseteq I$, and $\Phi_{j} \sse
\MTNF(X)$ for each $j \in J$. %
Note the restricted use of the conjunction symbol in disjunctive formulas:
a conjunction of the form $\bigwedge\Lambda\land\bigwedge_{j\in J}
\nab[j]\Phi_{j}$ will be called a \emph{special conjunction}.

\item
$\PNF(X)$ is the set of \emph{pure $\nb$-formulas} in $X$, generated by 
the following grammar:
\begin{align}
  \notag
  \phi & ::= \top \mid \bw\La \land \nbb \vec{\Phi}\,,
  \intertext{where $\La$ is a set of literals, $\vec{\Phi} =
    \set{\Phi_{i} \mid i \in I}$ is a vector such that, for each $i
    \in I$, $\Phi_{i}$ is a finite subset of $\PNF(X)$, and $\nbb
    \vec{\Phi}$ is defined by} %
  \label{eq:polynabla}
  \nbb \vec{\Phi} & : = \bigwedge_{i \in
    I} \nab[i]\Phi_{i}\,.
\end{align}
\end{enumerate}
\end{definition}

\begin{proposition}
\label{p:nb2}
\label{p:purenabla}
Let $X$ be a set of proposition letters.  
There are effective procedures 
\begin{enumerate}
\item
associating with each modal formula $\phi$ an equivalent $\nb$-formula;
\item
associating with each $\nb$-formula $\phi \in \MT(X)$ an equivalent
disjunctive formula;
\item
associating with each $\nb$-formula $\phi\in \MT(X)$ an equivalent
disjunction of pure $\nb$-formulas.
\end{enumerate}
\end{proposition}

\begin{proof}
Part~1 of the Proposition is proved by iteratively applying the equivalences
of \eqref{eq:nb0}, whereas part~2 is obtained by using \eqref{eq:nb1} as 
well as the distributive law of classical logic to push non special 
conjunctions to the leaves.  
For part~3, we first construct a formula $\phi' \in \MTNF(X)$ which is
equivalent to $\phi$. Using the fact that $\top$ is equivalent to
$\nab[i]\set{\top} \vee \nab[i]\nada$, we can suppose that, within
$\phi'$, each special conjunction $\bw \Lambda \land \bw_{j \in J}
\nab[j]\Phi_{j}$ is such that $J = I$. Then, we iteratively apply the
distributive law \eqref{eq:nb2} to $\phi'$ to push disjunctions up to the
root.
\end{proof}

Rewriting modal formulas into equivalent disjunctions of pure
$\nb$-formulas is not strictly necessary for our goals: we could work
with disjunctive formulas only.  However, we have chosen to consider
this further simplification because it drastically improves the
exposition of the next section.

\subsection*{Modal algebras and modal $\ff$-algebras}

We now move on to the algebraic perspective on flat modal fixpoint logic.
As usual in algebraic logic, formulas of the logic are considered as terms
over a signature whose function symbols are the logical connectives. 
Thus, from now on, the words ``term'' and ``formula'' will be considered
as synonyms.

Before we turn to the definition of the key concept, that of a modal
$\ff$-algebra, we briefly recall the definition of a modal algebra.
\begin{definition}
  Let $A = \struc{A,\bot,\top,\neg,\land,\lor}$ be a Boolean algebra.
  An operation $f: A \to A$ is called \emph{additive} if $f(a\lor b) =
  fa \lor fb$, \emph{normal} if $f\bot = \bot$, and an \emph{operator}
  if it is both additive and normal.  A \emph{modal algebra} (of type
  $I$) is a structure $A = \struc{A,\bot,\top,\neg,\land, \lor,\{
    \mop{i}^{A} \mid i \in I \}}$, such that the interpretation
  $\mop{i}^{A}$ of each action $i \in I$ is an operator on the Boolean
  algebra $\struc{A,\bot,\top,\neg, \land,\lor}$.
\end{definition}
Equivalently, a modal algebra is a Boolean algebra expanded with
operations that preserve all finite joins.  

Let $Z$ be a set of variables containing the free variables of a modal
formula $\phi$. If $A$ is a modal algebra, then $\phi^{A} : A^{Z} \rTo
A$ denotes the term function of $\phi$. Here $A^{Z}$ is the set of
$Z$-vectors (or $Z$-records), i.e. functions from the finite set $Z$
to $A$. Recall that if $\card(Z) = n$, then $A^{Z}$ is isomorphic to
the product of $A$ with itself $n$ times.
Next, given $\gamma \in \Gamma$, let us list its free variables as
usual, $\gamma = \ga(x,p_{1},\ldots,p_{n})$. Given a modal algebra $A$
the \emph{term function} of $\gamma$ is of the form $\ga^{A}: A \times
A^{n} \to A$.
Given a vector $\vec{b} = (b_{1},\ldots,b_{n}) \in A^{n}$, we let
$\ga^{A}_{\vec{b}}: A \to A$ denote the map given by
\begin{equation}
\label{eq:2a}
\ga^{A}_{\vec{b}}(a) := \ga^{A}(a,\vec{b})\,.
\end{equation}

\begin{definition}
A \emph{modal $\ff$-algebra} is a modal algebra $A$ endowed with an operation
$\ff_{\ga}^{A}$ for each $\ga\in\Ga$ such that for each $\vec{b}$, 
$\ff_{\ga}^{A}(\vec{b})$ is the least fixpoint of 
$\ga^{A}_{\vec{b}}$ as defined in (\ref{eq:2a}).
\end{definition}

Note that modal $\ff$-algebras are generally not complete; the definition
simply \emph{stipulates} that the least fixpoint exists, but there is no
reason to assume that this fixpoint is reached by ordinal approximations.

Recall that $f: A \rTo B$ is a modal algebra morphism if the operations
$\langle \bot,\top,\neg\linebreak \land, \set{\pos[i]\mid i \in I}\rangle$
are preserved by $f$. If $A$ and $B$ are also modal
$\ff$-algebras then $f$ is a \emph{modal $\ff$-algebra morphism}
if moreover each $\ff_{\gamma}$, $\gamma \in \Gamma$, is preserved by
$f$. 
This means that
\begin{align*}
  f(\ff^{A}_{\gamma}(\vec{v})) & =
  \ff^{B}_{\gamma}(f\circ\vec{v})\,,
\end{align*}
for each $\vec{v} \in A^{n}$ and $\gamma \in \Gamma$.  A
$\ff$-algebra morphism is an embedding if it is injective, and we say
that $A$ embeds into $B$ if there exists an embedding $f : A \rTo B$.

In this paper we will be mainly interested in two kinds of modal
$\ff$-algebras: the ``concrete'' or ``semantic'' ones that encode a
Kripke frame, and the ``axiomatic'' ones that can be seen as algebraic
versions of the axiom system $\Klogfxp$ to be defined in the next
section.  We first consider the concrete ones.  

\begin{definition}
Let $\bbS = \langle S,\set{R_{i}\mid i \in I} \rangle$ be a transition system.
Define, for each $i \in I$, the operation $\exop{R_{i}}$ by putting, for each
$X \subseteq S$,
$\exop{R_{i}}X = \set{ y \in S \mid \exists x \in X \tst y R_{i} x}$.
The \emph{$\ff$-complex algebra} is given as the structure 
\[
\bbS^{\ff} :=
\struc{\power(S),\nada,S,\mycomplement{(\,\cdot\,)},\cup,\cap,
\set{\exop{R_{i}}\mid i \in I}}.
\]
We will also call these structures \emph{Kripke $\ff$-algebras}.
\end{definition}

\begin{definition}
  \label{def:constructive}
  Let $A = \struc{A,\leq}$ be a partial order with least element
  $\bot$, and let $f: A \to A$ be an order-preserving map on $A$.  For
  $k \in \om$ and $a \in A$, we inductively define $f^{k}a$ by putting
  $f^{0}a := a$ and $f^{k+1}a := f(f^{k}a)$.  If $f$ has a least
  fixpoint $\mu.f$, then we say that this least fixpoint is
  \emph{constructive} if $\mu.f = \bigvee_{k\in\om}f^{k}(\bot)$.  
A modal $\ff$-algebra is called \emph{constructive} if $\ff_{\ga}^{A}(\vec{b})$
is a constructive least fixpoint, for each $\ga \in \Ga$ and each $\vec{b}$ in
$A$.
\end{definition}

\begin{remark}
Our terminology slightly deviates from that in~\cite{sant:comp08}, where the 
least fixpoint of an order-preserving map on a partial order is called 
constructive if it is equal to the join of all its ordinal approximations,
not just of the $\om$ first ones.
\end{remark}

\subsection*{$\O$-adjoints and fixpoints}

We now recall the
well known concept of adjointness, and briefly discuss its
generalization, $\O$-adjointness.
\begin{definition}
\label{d:adj}
Let $A = (A,\leq)$ and $B = (B,\leq)$ be two partial orders.
Suppose that $f: A\to B$ and $g: B \to A$ are order-preserving maps such that 
\begin{equation}
\label{eq:adj1}
fa \leq b \mbox{ iff } a \leq gb,
\end{equation}
for all $a \in A$ and $b \in B$.  
Then we call $(f,g)$ an \emph{adjoint pair}, and say that $f$ is the
\emph{left adjoint} of, or
  \emph{residuated} by, $g$, and that $g$ is the \emph{right adjoint},
  or residual, of $f$.
  We say that $f$ is an \emph{$\O$-adjoint} if it satisfies the weaker
  property that for every $b \in B$ there is a finite set $G_{f}(b)
  \sse A$ such that
  \[
  fa \leq b \mbox{ iff } a \leq a' \mbox{ for some } a' \in G_{f}(b),
  \]
  for all $a \in A$ and $b \in B$.
\end{definition}

\newcommand{\T}{\mathcal{T}}
\begin{remark}
  The \emph{terminology} `$\O$-adjoint' can be explained as follows.
  Let $\T$ be a functor on the category of partial orders (with
  order-preserving maps as arrows).  Call a morphism $f: (A,\leq) \rTo
  (B,\leq)$ a \emph{left $\T$-adjoint} if the map $\T f: \T(A,\leq)
  \rTo \T(B,\leq)$ has a right adjoint $G: \T(A,\leq) \rTo \T(B,\leq)$
  in the sense of (\ref{eq:adj1}) above.  Let now $\T$ be the functor
  $\Of$ defined as follows.
On objects, $\Of$ maps a partial order $(A,\leq)$ to the set $\Of(A,\leq)$
of finitely generated downsets of $(A,\leq)$, ordered by inclusion.
Alternatively, $\Of(A,\leq)$ is the free join-semilattice generated by
$(A,\leq)$.
To become a functor, $\Of$ takes an arrow $f: (A,\leq) \rTo (B,\leq)$ to
the function $\Of(f)$ that maps a subset $X \in \Of(A)$ to the set of points
that are below some element of the direct image $f(X)$.

We leave it as an exercise for the reader to verify that an order-preserving
  map $f$ is an $\O$-adjoint, in the sense of Definition~\ref{d:adj}
  iff it is a left $\Of$-adjoint in the sense just described.  We
  write $\O$-adjoint rather than left $\Of$-adjoint in order to keep
  our notation simple.

  \label{rem:quasiorder}
  Finally, observe that to define adjoints, $\T$-adjoints, and
  $\O$-adjoints, we do not need the antisymmetric law of partial
  order, we can define these notions for quasi orders.
\end{remark}

It is well known that left adjoint maps preserve all existing joins of
a poset.  Similarly, one may prove that $\O$-adjoints preserve all
existing joins of \emph{directed} sets.

$\O$-adjoints are relevant for the theory of least fixpoints because of the
following.
If $f : A^{n} \rTo A$ is an $\O$-adjoint, say that
$V \subseteq A$ is $f$-closed if $y \in V$ and $\vec{a} =
(a_{1},\ldots,a_{n}) \in G_{f}(y)$
implies $a_{i} \in V$ for $i = 1,\ldots ,n$.
If ${\cal F}$ is a family of $\O$-adjoints of the form $f : A^{n} \rTo A$, say that $V$
is $\mathcal{F}$-closed if it is $f$-closed for each $f \in \mathcal{F}$.
\begin{definition}
  \label{def:finitaryOadj}
  A family of $\O$-adjoints ${\cal F} = \set{f_{i}:A^{n_{i}}\rTo A\mid
    i \in I}$ is said to be \emph{finitary} if, for each $x \in A$,
  the least set ${\cal F}$-closed set containing $x$ is finite.  The
  $\O$-adjoint $f^{n} : A \rTo A$ is finitary if the singleton $\set{f}$
  is finitary.
\end{definition}
Clearly, if $f$ belongs to a finitary family, then it is finitary.
\begin{proposition}
  \label{p:foadjconstructive}
  If $f : A \rTo A$ is a finitary $\O$-adjoint, then its least
  prefixpoint, whenever it exists, is constructive.
\end{proposition}
See \cite[Proposition 6.6]{sant:comp08} for a proof of the Proposition.

\smallskip

The next Proposition collects the main properties of finitary families of
$\O$-adjoints. Roughly speaking, these properties assert that finitary
families may be supposed to be closed under composition, joining, and tupling.  
\begin{proposition}
  \label{prop:propertiesoadjoints}
  Let $\mathcal{F}$ be a finitary family of $\O$-adjoints on a modal
  algebra $A$. Suppose also that $f,g \in \mathcal{F}$. Then
  $\mathcal{G}$ also is a finitary family of $\O$-adjoints, whenever
  \begin{enumerate}
  \item $\mathcal{G} \subseteq \mathcal{F}$,
  \item $\mathcal{G} = \mathcal{F} \cup \set{h}$,  $f : A
    \times A^{Z} \rTo A$, $g : A^{Y} \rTo A$, and $h = f \circ (g
    \times A^{Z}) : A^{Y} \times A^{Z} \rTo A$, 
  \item $\mathcal{G} = \mathcal{F} \cup \set{h}$,  $f,g : A^{Z}
    \rTo A$, and $h = f \vee g$,  
  \item $\mathcal{G} = \set{F : A^{Z} \rTo A^{Z}}$ and $\set{\pi_{z}
      \circ F : A^{Z} \rTo A \mid z \in Z} \subseteq \mathcal{F}$.
  \end{enumerate}
\end{proposition}
\begin{proof}
Part~1 of the statement is obvious. For the parts~2~and~4, we invite the
reader to consult \cite[Lemmas 6.10 to 6.12]{sant:comp08}. For Part~3,
  observe that
  \begin{align*}
    G_{f \vee g}(d) & = G_{f}(d) \land G_{f}(d)\,, 
  \end{align*}
  where $C \land D = \set{\vec{v} \land \vec{u} \mid \vec{v} \in C
    \tand \vec{u}\in D}$. Thus, if $v_{0} \in A$ and $V$ is a finite
  $\mathcal{F}$-closed set with $v_{0} \in V$, then $V_{\land}$, the
  closure of $V$ under meets, is a finite $\mathcal{G}$-closed set
  with $v_{0} \in V_{\land}$.
\end{proof}

\subsection*{Systems of equations}

\begin{definition}
A \emph{modal system} or \emph{system of equations} is a pair $T = \langle Z,
\set{t_{z}}_{z \in Z} \rangle$ where $Z$ is a finite set of variables and
$t_{z} \in \MT(Z \cup P)$ for each $z \in Z$.  
Such a modal system is \emph{pointed} if it comes with a specified variable
$z_{0} \in Z$. 
\end{definition}

Given a modal system $T$ and a modal algebra $A$, there exists a
unique function %
$T^{A} : A^{Z} \times A^{P} \rTo A^{Z}$ %
such that, for each projection $\pi_{z} : A^{Z} \rTo A$, $\pi_{z}
\circ T^{A} = t_{z}^{A}$. We shall say that $T^{A}$ is the
interpretation of $T$ in $A$. Whenever it exists, we shall denote the
least fixpoint of $T^{A}$ by $\mu_{Z}.T^{A} : A^{P} \rTo A^{Z}$. %

In this paper we will be interested in modal systems where every term
is in a special syntactic shape.

\begin{definition}
In the monomodal setting, a term $t \in \MT(Z \cup P)$ is \emph{semi-simple}
if it is a disjunction of terms of the form $\La\land\nb\Phi$, where $\La$
is a set of $P$-literals, and each $\phi\in\Phi$ is a finite conjunction of
variables in $Z$ (where $\top$ is the empty conjunction).
For such a term to be \emph{simple}, we require that each $\phi\in\Phi$
belongs to the set $Z \cup \{\top\}$.  
In the polymodal setting, a term $t$ is semi-simple (simple) if it is a
disjunction of terms of the form $\La\land \bw_{j\in J}\nab[j]\Phi_{j}$,
where $J\sse I$ and each of the formulas in $\bigcup_{j}\Phi_{j}$ satisfies
the respective above-mentioned condition.
  
A  modal system $T = \langle Z,\set{t_{z}}_{z \in Z} \rangle$ is 
\emph{semi-simple} (\emph{simple}, respectively), if every term $t_{z}$ is 
semi-simple (simple, respectively).
\end{definition}


\section{The axiomatization $\Klogfxp(\Ga)$}
\label{s:ax}

\newcommand{\np}{\overline{p}}
\newcommand{\nq}{\overline{q}}

The axiom system $\Klogfxp(\Ga)$ that we will define in this section
adds, for each $\ga \in \Ga$, a number of axioms and derivation rules
to the basic (poly-)modal logic $\Klog$.  
We obtain these axioms and rules effectively, via some systems of equations
that we will associate with $\ga$.  Here is a summary of the procedure.

\begin{enumerate}
\item[0.] \emph{Preprocess}, rewriting $\ga(x)$ as a guarded
  disjunction of special pure $\nb$-formulas.
\item \emph{Represent} each such $\ga$ by a \emph{semi-simple} system
  of equations $T_{\ga}$.
\item \emph{Simulate} $T_{\ga}$ by a \emph{simple} system of equations
  $T_{\ga}^{+}$.
\item \emph{Read off} the axiomatization for $\ff_{\ga}$ from
  $T_{\ga}^{+}$.
\end{enumerate}

The aim of this section is to define and discuss this procedure in
full detail --- readers who only want to look at the definition of the
axiom system can proceed directly via the Definitions~\ref{d:Tgamma},
\ref{d:sim} and~\ref{d:axKlogfx}.
For the sake of readability we work mainly in the monomodal framework.

Before carrying on, let us fix some notation to be used throughout
this section.
We shall use the capital letters $X,Y,Z$ to
denote sets of fixpoint variables. On the other hand, $P$ will denote
a set of proposition letters not containing any of these fixpoint
variables.
If $\tau \in \MT(X\cup P)$ and $\set{\si_{y} \mid y \in Y} \subseteq
\MT(X)$ is a collection of terms indexed by $Y\sse X$, then we shall denote
by $\vec{\si}$ such a collection, and by $\tau[\vec{\si}/\vec{y}]$ the
result of simultaneously substituting every variable $y \in Y$ with the
term $\si_{y}$.

\subsection*{Preprocessing $\ga$}

Fix a modal formula $\ga(x)$ in which the variable $x$ occurs only
positively.  First of all, for our purposes we may assume that each
occurrence of $x$ is \emph{guarded} in $\ga$, that is, within the
scope of some modal operator.  In the theory of fixpoint logics it is
well-known that this assumption is without loss of generality, see for
example \cite[Proposition 2]{walukiewicz}. In order to give a quick
justification, recall that our goal is to axiomatize the least
prefixpoint of $\ga(x)$.  If $x$ is not guarded in $\ga$, then we
can find terms $\ga_{1},\ga_{2}$, with $x$ guarded in both $\ga_{1}$ and
$\ga_{2}$, and such that the equation
\begin{align*}
  \ga(x,\vec{p}) & = (x \land \ga_{1}(x,\vec{p})) \vee
  \ga_{2}(x,\vec{p}) \,,
\end{align*}
holds on every modal algebra.
It is easily seen that, on every modal algebra, $\ga$ and $\ga_{2}$ have
the same set of prefixpoints.  
Thus, instead of axiomatizing $\ff_{\ga}$, we can equivalently axiomatize
$\ff_{\ga_{2}}$.

Second, given the results mentioned in the previous section, we may
assume that $\ga$ is a disjunction of pure $\nb$-formulas
(cf.~Proposition~\ref{p:purenabla}).  However,
given the special role of the variable $x$, it will be convenient for
us to modify our notation accordingly.  We introduce the following
abbreviation:
\[
\nab[\La]\Phi := \bigwedge\La\land\nb\Phi,
\]
in the case that $\La \sse \Lit(X)$ and $x$ does not occur in $\La$.

\begin{definition}
  Given a set $P$ of proposition letters and a variable $x\not\in P$,
  we define the set
  of 
  \emph{pure $\nbx$-formulas in $P$} by the following grammar:
  \begin{equation}
    \label{eq:ga0}
    \phi ::=  
    \top \mid x \mid
    \nab[\La]\Phi \mid x \land \nab[\La]\Phi,
  \end{equation}
  where $\La \sse \Lit(P)$, and $\Phi$ is a set of pure $\nbx$-formulas in
  $P$.
\end{definition}

\begin{remark}
\label{rem:defofvecnab}
Recall from equation~\ref{eq:polynabla} that, in the polymodal setting, $\nbb\vec{\Phi}$ denotes the
formula $\bw_{i\in I} \nab[i]\Phi_{i}$, where $\vec{\Phi}$ is the
vector $\{ \Phi_{i} \mid i \in I \}$.  Now we can define the set of
$\nbbx$-formulas in $P$, in the polymodal setting, by the following grammar:
\[
\phi ::=  
\top \mid x \mid
\nabb[\La]\vec{\Phi} \mid x \land \nabb[\La]\vec{\Phi}\,.
\]
Then basically, the algorithm for obtaining the axiomatization in the
polymodal case works the same as in the monomodal case, with the 
polymodal %
nabla-operator $\nbb$ replacing the monomodal $\nb$.
\end{remark}

\begin{convention}
In concrete examples we will denote the set $\La$ in $\nab[\La]$ as a
\emph{list} rather than as a set, and write $\np$ rather than $\neg p$.
For instance we will write $\nab[p\nq]\Phi$ instead of
  $\nab[\{p,\neg q\}]\Phi$.  Furthermore, we will write $\nb\Phi$
  instead of $\nab[\nada]\Phi$.
\end{convention}

\begin{lemma}
\label{l:pnf}
Every modal formula $\ga \in \MT(P \cup \set{x})$ in which the
variable $x$ only occurs positively can be effectively rewritten as an
equivalent disjunction $\ga'$ of pure $\nbx$-formulas in $P$.  Furthermore,
if $x$ is guarded in $\ga$ then $x$ is guarded in $\ga'$ as well.
\end{lemma}

\begin{proof}
  In Proposition \ref{p:purenabla} we saw that every modal formula
  $\ga$ can be equivalently rewritten as a disjunction $\ga'$ of pure
  $\nb$-formulas.  If $x$ occurs only positively in $\ga$, then this
  formula will have no subformulas of the form $\bw\La \land \nb\Phi$
  with $\neg x \in \La$.  From this the lemma is immediate.
\end{proof}

\begin{example}
\label{ex:1}
Consider the formula $(p \land \Box x) \lor (\neg p \land \dia (x
\land \dia x))$.
Rewriting this as a disjunction of pure $\nbx$-formulas, we obtain
\begin{equation}
\label{eq:ex1}
\ga(x) = \nab[p]\nada \lor \nab[p]\{x\} \lor 
\nab[\np]\set{ \top, x \land \nb\set{\top,x} }\,.
\end{equation}
\end{example}

\subsection*{Step 1: from formulas to semi-simple systems of equations}

In the first step of the procedure, we represent a formula $\ga$ as a 
semi-simple system of equations $T_{\ga}$.
Fix a modal formula $\ga(x)$ in which the variable $x$ only occurs positively.
Without loss of generality we may assume that $\ga$ is a disjunction of 
pure $\nbx$-formulas, and guarded in $x$.
Roughly speaking, to obtain the modal system $T_{\ga}$ we cut up the formula
$\ga$ in layers, step by step peeling off its modalities and introducing new
variables for (some of) $\ga$'s subformulas of the form $\nab[\La]\Phi$.

\begin{definition}
Let $\ga(x) \in \MT(P \cup \set{x})$ be a disjunction of pure $\nbx$-formulas,
and guarded in $x$.
We define $\SC_{\ga}$, the set of \emph{special conjunctions} in $\ga$, as the
set of subformulas of $\ga$ of the form $\nab[\La]\Phi$.
$\SC'_{\ga}$ is
the set of special conjunctions that occur in the scope of some $\nb$-formula.
Furthermore, we define $\RSF{\ga} := \set{\ga} \cup \SC'_{\ga}$ as the set of
\emph{relevant subformulas} of $\ga$.
\end{definition}

To see the difference between the sets $\SC'_{\ga}$ and $\SC_{\ga}$, observe
that $\ga$ itself is a disjunction of special conjunctions.
These disjuncts are elements of $\SC_{\ga}$, but we only put them in 
$\SC'_{\ga}$ if they occur as subformulas of $\ga$ deeper in the formula tree
as well.

\begin{example}
\label{ex:2}
With $\ga$ the formula given by (\ref{eq:ex1}), we find that $\SC_{\ga}$
consists of the four formulas
\begin{eqnarray*}
\psi_{1} & = & \nab[p]\nada,
\\ \psi_{2} & = & \nab[p]\{x\}, 
\\ \psi_{3} & = & \nab[\np]\{ \top, x \land \nb\{\top,x\} \},
\\ \psi_{4} & = & \nb\{\top,x\}.
\end{eqnarray*}
Of these, only $\psi_{4}$ makes it into $\SC'_{\ga}$, so $\RSF{\ga} = \{ \ga,
\psi_{4} \}$.
\end{example}

The system of equations $T_{\ga}$ will be based on a set of variables that is
in one-to-one correspondence with the set of relevant formulas.

\begin{definition}
Let $\ga(x) \in \MT(P \cup \set{x})$ be a disjunction of pure $\nbx$-formulas,
and guarded in $x$.
Let 
\[
Z = \set{ z_{\psi}
  \mid \psi \in \RSF{\ga} }
\]
be a set of fresh variables (in one-to-one correspondence with the set
$\RSF{\ga}$), and let $[\vec{\psi}/\vec{z}]$ be the natural
substitution replacing each variable $z_{\psi}$ with the formula $\psi$.
\end{definition}

The key observation in the definition of the modal system $T_{\ga}$ is that
every disjunction of formulas in $\SC_{\ga}$ can be seen as the
$[\vec{\psi}/\vec{z}]$-substitution instance of a semi-simple formula
$\widehat{\psi}$.
For instance, in Example~\ref{ex:2}, writing 
\[
\widehat{\psi_{3}} = \nab[\np] \{ \top, x \land z_{\psi_{4}} \},
\]
we have that $\psi_{3} = \widehat{\psi_{3}}[\psi_{4}/z_{\psi_{4}}]$.

\begin{lemma}
For every formula $\psi \in \RSF{\ga}$ there is a 
semi-simple formula $\widehat{\psi}$ such that 
$\psi = \widehat{\psi}[\vec{\psi}/\vec{z}]$.
\end{lemma}

\begin{proof}
Given a special conjunction $\nab[\La]\Phi$ in $\ga$, each $\phi\in\Phi$
has one of the forms $\top,x,\psi$, or $x \land\psi$, where $\psi$ is
again a special conjunction.
Let $\widehat{\nab[\La]\Phi}$ be the formula we obtain by replacing $\Phi$'s
elements of the form $\psi$ and $x\land\psi$ with $z_{\psi}$ and $x\land
z_{\psi}$, respectively.
It is immediate that $\nab[\La]\Phi =
\widehat{\nab[\La]\Phi}[\vec{\psi}/\vec{z}]$.
This takes care of the formulas $\psi \in \SC'_{\ga}$, while for $\ga$, which
can be written as a disjunction $\bigvee_{i}\phi_{i}$ of special conjunctions,
we can simply take the formula $\widehat{\ga} :=
\bigvee_{i}\widehat{\phi_{i}}$.
It is easy to see that the obtained formulas are semi-simple.
\end{proof}

\begin{definition}
\label{d:Tgamma}
Let $\ga(x) \in \MT(P \cup \set{x})$ be a disjunction of pure $\nbx$-formulas,
and guarded in $x$.
For $z = z_{\psi} \in Z$, we write $\rho_{z} := \widehat{\psi}$, and let
$\tau_{z}$ denote the term $\rho_{z}[z_{\ga}/x]$.
We call the modal system
\[
T_{\ga} := \langle Z, \set{\tau_{z}\mid z \in Z} \rangle
\]
the \emph{system representation of $\ga$}.
$T_{\ga}$ is pointed by the variable $z_{\ga}$. 
\end{definition}

The reader will have no difficulties verifying that $T_{\ga}$ is a semi-simple
systems of equations.

\begin{example}
\label{ex:3}
For the formula $\ga$ of the Examples~\ref{ex:1}/\ref{ex:2}, we obtain
(writing $z_{i}$ rather than $z_{\psi_{i}}$) the following system $T_{\ga}$.
As its variables it has the set $\set{z_{\ga},z_{4}}$, and its equations are
the following:
\[\begin{array}{lll}
z_{\ga} &=& 
\nab[p]\nada \lor \nab[p]\{z_{\ga}\} \lor 
\nab[\np] \{ \top, z_{\ga} \land z_{4} \}
\\ z_{4} &=& \nb\{\top,z_{\ga}\}.
\end{array}\]
\end{example}

We call the modal system $T_{\ga}$ a \emph{representation} of the formula $\ga$
because the least fixpoints of $T_{\ga}$ and $\ga$ are mutually expressible
--- for the precise formulation of this statement we refer to
Proposition~\ref{p:lfpsoe1} below.
Here we just mention the key observation underlying this proposition, which
relates the (parametrized) fixpoints of $T_{\ga}$ to those of $\ga$, as
follows.

\begin{proposition}
  \label{p:fpsoe1}
  Let $\ga$ be a modal formula in which the variable $x$ only occurs
  positively, let $A$ be a modal algebra, and $\vec{v} \in A^{P}$ a
  sequence of parameters in $A$.
  \begin{enumerate}
  \item
    If $a \in A$ is a fixpoint of $\ga^{A}_{\vec{v}}$, then the vector
    $\set{ \psi^{A}(a,\vec{v}) \mid \psi \in \RSF{\ga} }$
    is a fixpoint of $(T_{\ga}^{A})_{\vec{v}}$.
  \item
    If $\set{ b_{\psi} \mid \psi \in \RSF{\ga} }$ is a fixpoint of
    $(T_{\ga}^{A})_{\vec{v}}$, then $b_{\ga} \in A$ is a fixpoint of
    $\ga^{A}_{\vec{v}}$.
  \end{enumerate}
\end{proposition}

\begin{proof}
Immediate by the definitions.
\end{proof}

Since our main aim is to represent $\ga$ by a \emph{simple} set of
equations, formulas $\ga$ for which $T_{\ga}$ itself is already
simple, are clearly of interest. We shall introduce in
Section~\ref{s:sc} classes of formulas, called \emph{untied} and
\emph{harmless}, that have this property. If every formula $\gamma \in
\Gamma$ belongs to those classes, then we can prove that
$\Klogfx(\Gamma)$ is already  a complete and sound axiom system.

\subsection*{Step 2: from semi-simple systems of equations to simple ones}

The second step of our procedure is based on the \emph{subset
  construction} of Arnold \& Niwi\'{n}ski~\cite{arnoldniwinski}.  The
idea behind this construction is that, under some conditions, one may
eliminate conjunctions from a system of equations $T$ through
\emph{simulating} it by another system, $T^{+}$.  Roughly, the idea of
the construction is that the variables of the system $T^{+}$
correspond to the conjunctions of the non-empty sets of variables of
the system $T$.

\begin{convention}
Given the set of variables $Z$, we let $Y = \set{y_{S} \mid S \in \Pplus(Z)}$
be a set of new variables in bijection with $\Pplus(Z)$, the set of non empty
subsets of $Z$.
For $S \in \Pplus(Z)$, we denote by $z_{S}$ the term
$\bigwedge S$,
and let $[\vec{z}/\vec{y}]$ denote the substitution which replaces each
variable $y = y_{S} \in Y$ with the term $z_{S}$.
\end{convention}

The following lemma is the heart of the simulation construction.

\begin{proposition}
\label{p:simterm}
Let $\set{\tau_{i} \mid i \in I}$ be a finite collection of semi-simple terms
in $Z$.
\begin{enumerate}
\item
There is a semi-simple term $\tau$ in $Z$ which is equivalent to 
$\bigwedge_{i\in I} \tau_{i}$.
\item
There is a simple term $\si$ in $Y$, such that the term $\si[\vec{z}/\vec{y}]$
is equivalent to $\bigwedge_{i\in I} \tau_{i}$.
\end{enumerate}
\end{proposition}

\begin{proof}
  We give the proof in the monomodal setting.  
  The first part of the lemma
  follows easily from successive applications of the distributive law
  (\ref{eq:nb1}) for the cover modality.  Obviously it suffices to
  prove that the conjunction of two semi-simple terms $\bw\La \land
  \nb\Phi$ and $\bw\La \land \nb\Phi'$ is semi-simple.  But by
  (\ref{eq:nb1}), and the distributive law of classical propositional
  logic, this conjunction is equivalent to some formula $\bw (\La \cup
  \La') \land \nb\Psi$, where each formula $\psi \in \Psi$ is of the
form $\phi\land\phi'$, with $\phi\in\Phi$ and $\phi'\in\Phi'$, and thus
itself a finite conjunction of variables in $Z$.
In other words, the formula $\bw (\La \cup \La') \land \nb\Psi$ is equivalent
to a semi-simple formula.

  The second part of the proposition is an almost immediate
  consequence of the first, by the observation that with every
  semi-simple term $\tau$ we may associate a simple term $\si$ such
  that $\tau$ is equivalent to the term $\si[\vec{z}/\vec{y}]$.  
The term $\si$ is obtained from $\tau$ simply by replacing, for each
disjunct $\La \land \nb\Phi$, each formula $\bigwedge S \in \Phi$ (with
$S\neq\nada$) by the variable $y_{S}$.
\end{proof}

\begin{remark}
It should be immediate to see how modify the above proof for the
  setting of polymodal logic. Indeed, recall first from
  Remark~\ref{rem:defofvecnab} the definition of the polymodal
  $\nbb$. Trivially, one has
  \begin{align*}
    \bw \Lambda \land \nbb \vec \Phi 
    \land \bw \Lambda'  \land \nbb \vec \Psi
    & = \bw (\Lambda \cup \Lambda')
    \land \bw_{i \in I} \nab[i] \Phi_{i} \land \nab[i] \Psi_{i}\,, 
  \end{align*}
  so that, by applying first the laws (\ref{eq:nb1}) for each
  $\nab[i]$, and then the distributive law of classical propositional
  logic, a fundamental distributive law for the polymodal $\nbb$ may
  also be derived.
\end{remark}

\begin{definition}
\label{d:sim}
Let $T = \langle Z, \{ \tau_{z} \mid z \in Z \}\rangle$ be a semi-simple
modal system.
For any $y \in Y$, writing $y = y_{S}$ with $S \in \Pplus(Z)$, let $\si_{y}$ be
the simple term corresponding to
the conjunction $\bigwedge_{z\in S} \tau_{z}$, as provided by
Proposition~\ref{p:simterm}.
The \emph{simulation} of $T$ is defined as the system of equations
\[ T^{+} := \langle Y, \{ \si_{y} \mid y \in Y \} \rangle.\]
\end{definition}

\begin{example}
\label{ex:4}
Continuing Example~\ref{ex:3}, we may write
\[\begin{array}{lll}
  z_{\ga}\land z_{4} &=& 
  \big( \nab[p]\nada \land \nb\{\top,z_{\ga}\} \big)
  \lor
  \big( \nab[p]\{z_{\ga}\} \land \nb\{\top,z_{\ga}\} \big)
  \lor 
  \big( \nab[\np] \{ \top, z_{\ga} \land z_{4} \} \land \nb\{\top,z_{\ga}\} \big)
  \\ &=& \bot \lor \nab[p]\{z_{\ga}\} \lor 
  \nab[\np] \{ \top, z_{\ga} \land z_{4} ,z_{\ga}\} 
  \\ &=& \nab[p]\{z_{\ga}\}  \lor 
  \nab[\np] \{ \top, z_{\ga} \land z_{4} ,z_{\ga}\}\,,
\end{array}\]
where we have used some ``$\nb$-arithmetic'' to simplify the outcome.

Thus we obtain the following as the system $T_{\ga}^{+}$:
\[\begin{array}{lll}
  y_{\ga} &=& 
  \nab[p]\nada \lor \nab[p]\{y_{\ga}\} \lor 
  \nab[\np] \{ \top, y_{\ga4} \}
  \\ y_{4}    &=& \nb\{\top,y_{\ga}\}
  \\ y_{\ga4} &=& \nab[p]\{y_{\ga}\} \lor \nab[\np] \{ \top, y_{\ga4}, y_{\ga}\}\,. 
\end{array}\]
Here we write $y_{\ga}$ instead of $y_{\{\ga\}}$, etc.
\end{example}

For a more elaborate example, consider the following.

\begin{example}
Let $T$ be the semi-simple modal system given by
\[
\left\{\begin{array}{lll}
z_{1} &=& \nab[pq]\{ z_{1}\land z_{2}, z_{1}\land z_{3}\} \lor 
   \nab[p\nq]\{z_{2}\}\\
z_{2} &=& \nab[p]\{ z_{1}, z_{3} \}\\
z_{3} &=& \nab[]\{ z_{2}\land z_{3} \}\,.
\end{array}\right.
\]
Using the distributive laws for $\nb$ and some further $\nb$-arithmetic,
one may derive that
\[\begin{array}{lll}
z_{1}\land z_{2} 
&=& \nab[pq]\{ z_{1}\land z_{2}, z_{1}\land z_{3}\}
   \lor \nab[pq]\{ z_{1}\land z_{3}, z_{1}\land z_{2} \land z_{3} \}
\\ && \hspace*{10mm}
   \lor \nab[pq]\{ z_{1}\land z_{2}, z_{1}\land z_{3}, z_{1}\land z_{2} \land z_{3}\}
\\
z_{1}\land z_{3} 
&=& \nab[pq]\{ z_{1} \land z_{2} \land z_{3} \} 
   \lor \nab[\np q]\{z_{2} \land z_{3} \}
\\
z_{2}\land z_{3} 
&=& \nab[p]\{ z_{2}\land z_{3}, z_{1}\land z_{2} \land z_{3} \}
\\
z_{1}\land z_{2}\land z_{3} 
&=& \nab[pq]\{z_{1}\land z_{2}\land z_{3} \}\,.
\end{array}\]
From this it is easy to see that the simulation $T^{+}$ is given by
\[
\left\{\begin{array}{lll}
y_{1}  &=& \nab[pq]\{ y_{12}, y_{13}\} \lor \nab[p\nq]\{y_{2}\}\\
y_{2}  &=& \nab[p]\{ y_{1}, y_{3} \}\\
y_{3}  &=& \nab[]\{ y_{23} \}\\
y_{12} &=& \nab[pq]\{ y_{12}, y_{13}\} \lor \nab[pq]\{ y_{13}, y_{123} \}
    \lor \nab[pq]\{ y_{13}, y_{123} \}\\
y_{13} &=& \nab[pq]\{ y_{123} \} \lor \nab[\np q]\{y_{23} \}\\
y_{23} &=& \nab[p]\{ y_{23}, y_{123} \}\\
y_{123}&=& \nab[pq]\{ y_{123} \}\,,
\end{array}\right.
\]
where we write $y_{12}$ for $y_{\{1,2\}}$, etc.
\end{example}

The relation between the modal systems $T$ and $T^{+}$ is perhaps clarified
by a diagram. 
Let, for some modal algebra $A$, $\iota^{A}: A^{Z} \to A^{Y}$ be given by 
\begin{equation}
\label{eq:iota}
\iota^{A}(\vec{a})(y_{S})  = \bigwedge_{z \in S} a_{z}.
\end{equation}
Then Proposition~\ref{p:simterm}(2) maybe understood as stating that,
given a semi-simple system $T$, there exists a simple system $T^{+}$
such that, for every modal algebra $A$ and every parameter $\vec{v}
\in A^{P}$, the diagram
\begin{align}
  \label{diagram:powerset}
  &
  \xygraph{
    []*+{A^{Z}}="U" 
    (:[rrr]*{\,\,\,A^{Z}}="RU"^{T^{A}_{\vec{v}}}) 
    :[dd]*+{A^{Y}}^{\iota^{A}} 
    :[rrr]*+{A^{Y}}="RD"^{(T^{+})^{A}_{\vec{v}}}
    "RU":"RD"^{\iota^{A}}
  }
\end{align}
commutes.

On \emph{complete} modal algebras, the modal systems $T$ and $T^{+}$
are equivalent in the sense that the respective least fixpoints are
mutually definable --- this is in fact the point behind the
introduction of $T^{+}$ in~\cite{arnoldniwinski}.  In general however,
the relation between $T$ and $T^{+}$ seems to be less tight than that
between the formula $\ga$ (or rather, the system $\langle \set{x},
\set{\ga} \rangle$) and the system $T_{\ga}$.  In the next section we
discuss this relation in more detail: here we confine ourselves to the
following basic observation concerning fixpoints of $T$ and
$T_{\ga}$.

\begin{proposition}
  \label{p:fpsoe2}
  Let $T$ be a semi-simple modal system, let $A$ be a modal algebra,
  and $\vec{v} \in A^{P}$ a sequence of parameters in $A$.  If $\set{
    a_{z} \mid z \in Z}$ is a fixpoint of $T_{\vec{v}}$, then
  $\set{ \bigwedge \{ a_{z} \mid z\in S \} \mid S \in \Pplus(Z) }$ is
  a fixpoint of $T^{+}_{\vec{v}}$.
\end{proposition}

\begin{proof}
Immediate by (\ref{diagram:powerset}) and the definitions.
\end{proof}

\subsection*{Step 3: read off the axiomatization}

We are now ready to define the axioms and derivation rules that we
associate with a formula $\ga(x,\vec{p})$ in which the variable $x$
occurs only positively.  As we will see, these axioms and rules can be
easily read off from the simple modal system $T^{+}_{\ga}$ that we
obtained in the previous step of the procedure.  Before going into
the syntactic details, let us first take an algebraic perspective.

Let $A$ be a modal $\ff$-algebra, and let $\vec{v} \in A^{P}$ be a
sequence of parameters in $A$.  Since $\ff\vec{v}$ is the least fixpoint
of the map $\ga^{A}_{\vec{v}}: A \rTo A$, it follows from
Proposition~\ref{p:lfpsoe1} below that the vector 
\begin{equation}
\label{eq:crx1}
\big\{
\psi^{A}(\ff\vec{v}, \vec{v}) \mid \psi \in \RSF{\ga} \big\}
\end{equation}
is the least fixpoint of $(T^{A}_{\ga})_{\vec{v}}$. 
In order to arrive at a succinct presentation of our axiom system, it will
be convenient to think of the coordinate $\ga^{A}(\ff\vec{v},\vec{v})$ of
(\ref{eq:crx1}) (that is, the case where $\psi=\ga\in\RSF{\ga}$), as the
fixpoint $\ff\vec{v}$ itself --- this is allowed since $A$ is a modal
$\ff$-algebra.
For this purpose we introduce the following notation, using the one-to-one
correspondence between the sets $Z$ and $\RSF{\ga}$:
\[
\chi_{z} := \left\{\begin{array}{ll}
  x          & \mbox{if } \psi_{z} = \ga,
\\  \psi_{z} & \mbox{otherwise}.
\end{array}\right.
\]
We may conclude that on any modal $\ff$-algebra $A$, the set 
\begin{equation}
\label{eq:crx2}
\big\{
\chi_{z}^{A}(\ff\vec{v}, \vec{v}) \mid z \in Z \big\}
\end{equation}
is the least fixpoint of $(T^{A}_{\ga})_{\vec{v}}$. 
Then on the basis of Proposition~\ref{p:fpsoe2}, the set
\begin{equation}
\label{eq:crx3}
\Big\{ \;
\bigwedge_{z \in S} \chi_{z}^{A}(\ff\vec{v},\vec{v}) 
\;\mid\; S \in \Pplus (Z) 
\;\Big\}
\end{equation}
is \emph{some} fixpoint of $(T^{+}_{\ga})^{A}_{\vec{v}}$.
In case $A$ is a complete algebra, the results of Arnold \& Niwi\'{n}ski~%
\cite[\S 9]{arnoldniwinski} imply that (\ref{eq:crx3}) is in fact the
\emph{least} fixpoint of $(T^{+}_{\ga})^{A}_{\vec{v}}$.
For a general $\ff$-algebra, however, we have no justification for drawing
this conclusion.
This means that the following is a meaningful definition.

\begin{definition}
  A modal $\ff$-algebra $A$ is called \emph{regular} if for each
  $\ga\in\Ga$ and each $\vec{v} \in A^{P}$, the set (\ref{eq:crx3}) is
  the least fixpoint of $(T^{+}_{\ga})^{A}_{\vec{v}}$.
\end{definition}

We can now give an intuitive introduction of the axiom system
$\Klogfxp(\Ga)$ by saying that it expresses the regularity of modal
$\ff$-algebras.  In other words, our axiomatization requires that the
set (\ref{eq:crx3}) is the least fixpoint of
$(T^{+}_{\ga})^{A}_{\vec{v}}$.  Thus the above-mentioned result by
Arnold \& Niwi\'nski will imply the \emph{soundness} of the
axiomatization.

\begin{example}
\label{ex:5}
Continuing Example~\ref{ex:4}, we find that $\chi_{0} = x$ and $\chi_{4} =
\nb\{\top,x\}$.
Our axiomatization will express that, for any formula $\phi$ 
(corresponding to the sequence $\vec{v}$ of parameters), the vector
\[
\left(\begin{array}{c}
\ff \phi \\ \nb\{\top,\ff \phi\} \\ \ff \phi \land \nb\{\top,\ff \phi\}
\end{array}\right)
\;=\;
\left(\begin{array}{c}
\chi_{0}[\ff p/x][\phi/p] \\ \chi_{4}[\ff p/x][\phi/p] \\ 
(\chi_{0} \land \chi_{4})[\ff p/x][\phi/p]
\end{array}\right)
\]
is the least fixpoint of the system $T_{\ga}^{+}[\phi/p]$.
It suffices for our axiom system to express this for the proposition letter
$p$: a uniform substitution will then take care of the parameter $\phi$
(see footnote~\ref{fn:1} on how we formulate and interpret derivation 
rules).
Recall that the following $\si_{\ga}$, $\si_{4}$ and $\si_{\ga4}$ are the
terms of the system $T_{\ga}^{+}$:
\[\begin{array}{lll}
\si_{\ga} &=& 
   \nab[p]\nada \lor \nab[p]\{y_{\ga}\} \lor \nab[\np] \{ \top, y_{\ga4} \}
\\ \si_{4}    &=& \nb\{\top,y_{\ga}\}
\\ \si_{\ga4} &=& \nab[p]\{y_{\ga}\} \lor \nab[\np] \{ \top, y_{\ga4}, y_{\ga}\} \,.
\end{array}\]
Thus our axiomatization will contain the axioms
\begin{align}
  \label{ax:A0}%
  \tag{$A_{\ga}$}%
   \nab[p]\nada \lor \nab[p]\{\ff p\} \lor 
    \nab[\np] \{ \top,p \land \nb\{\top,\ff p \}\}
   & \to \ff p
   \\
   \label{ax:A4}%
   \tag{$A_{4}$}%
    \nb\{\top,\ff p\}
   & \to \nb\{\top,\ff p\}
   \\
   \label{ax:A04}%
   \tag{$A_{\ga4}$}%
   \nab[p]\{ \ff p\} \lor \nab[\np] \{ \top, p \land \nb\{\top,\ff p,
   \ff p\} & \to p \land \nb\{\top,\ff p \} &
\end{align}
stating that
$\left(\begin{array}{c}
\ff p  \\ \nb\{\top,\ff p \} \\ \ff p  \land \nb\{\top,\ff p \}
\end{array}\right)$
is a prefixpoint of the system $T_{\ga}^{+}$, and the derivation rules
\begin{align}
\label{rule:R0}%
\tag{$R_{\ga}$}%
\frac{\nab[p]\nada \lor \nab[p]\{y_{\ga}\} \lor \nab[\np] \{ \top, y_{\ga4} \}
          \to y_{\ga} \;\;\;\; 
   \nb\{\top,y_{\ga}\}\to y_{4} \;\;\;\; 
   \nab[p]\{y_{\ga}\} \lor \nab[\np] \{ \top, y_{\ga4}, y_{\ga}\} \to y_{\ga4} }
 {\ff p \to y_{\ga}}
\\[4mm]
\label{rule:R4}%
\tag{$R_{4}$}%
\frac{\nab[p]\nada \lor \nab[p]\{y_{\ga}\} \lor \nab[\np] \{ \top, y_{\ga4} \}
         \to y_{\ga} \;\;\;\; 
   \nb\{\top,y_{\ga}\}\to y_{4} \;\;\;\; 
   \nab[p]\{y_{\ga}\} \lor \nab[\np] \{ \top, y_{\ga4}, y_{\ga}\} \to y_{\ga4} }
 {\nb\{\top,\ff p\} \to y_{4}}
\\[4mm]
\label{rule:R04}%
\tag{$R_{\ga4}$}%
\frac{\nab[p]\nada \lor \nab[p]\{y_{\ga}\} \lor \nab[\np] \{ \top, y_{\ga4} \}
         \to y_{\ga} \;\;\;\; 
   \nb\{\top,y_{\ga}\}\to y_{4} \;\;\;\; 
   \nab[p]\{y_{\ga}\} \lor \nab[\np] \{ \top, y_{\ga4}, y_{\ga}\} \to y_{\ga4} }
 {p \land \nb\{\top,\ff p \} \to y_{\ga4}}
\end{align}
expressing that this same vector is the \emph{least} of the prefixpoints of
$T_{\ga}^{+}$.
\end{example}

In order to address the general case, we discuss some notational issues.
Given $S \in \Pplus(Z)$, let $\chi^{\ff}_{S}$ denote the 
following formula
\begin{align*}
\chi^{\ff}_{S} & = \bigwedge_{z \in S} \chi_{z}[\ff_{\ga}/x],
\end{align*}
and, as usual, let $\vec{\chi^{\ff}}$ be the vector of terms $\big\{
\chi^{\ff}_{S} \mid S \in \Pplus(Z) \big\}$. 
Using the one-to-one correspondence between the sets $Y$ and $\Pplus(Z)$,
we let $[\vec{\chi^{\ff}}/\vec{y}]$ denote the substitution which replaces
each variable $y = y_{S}$ with the formula $\chi^{\ff}_{S}$.  Furthermore,
recall that $\set{ \si_{S} \mid S \in \Pplus(Z)}$ is the vector of terms
of the modal system $T_{\ga}^{+}$.

\begin{definition}
\label{d:axKlogfx}
The axiom system $\Klogfxp(\ga)$ is obtained by adding to the axiomatization
$\Klogfx(\ga)$ of Definition~\ref{d:ax1}, for each $S \in \Pplus(Z)$, the
following axiom:
\begin{align*}
  \label{ax:AS}%
  \tag{$A_{S}$}%
&\si_{S}[ \vec{\chi^{\ff}}/\vec{y}] \to  \chi^{\ff}_{S}\,, 
\intertext{%
as well as the following derivation rule:%
}
  \label{rule:RS}%
  \tag{$R_{S}$}%
& \frac{ \{\; \si_{Q} \to y_{Q} \;\mid\; Q \in \Pplus(Z) \} 
     }{ \chi^{\ff}_{S} \to y_{S} }
\end{align*}

Finally, the axiom system $\Klogfxp(\Ga)$ is obtained as the union of all the
axioms and inference rules of the axiom systems $\Klogfxp(\ga)$, $\ga\in\Ga$.
\end{definition}

\begin{remark}
Strictly speaking, we no longer need the axiom \eqref{ax:sharpprefix} and
the rule \eqref{ax:sharpleast} since it can be proved on the basis of 
Proposition~\ref{p:lfpsoe2} and the results in Section~\ref{sec:constructive},
that \eqref{ax:sharpprefix} is derivable and that \eqref{ax:sharpleast} is
admissible in the system obtained by deleting \eqref{ax:sharpprefix} and
\eqref{ax:sharpleast} from $\Klogfxp(\Ga)$.
\end{remark}

\begin{remark}
It is not hard to see that the number of rules and axioms that we add to 
$\Klogfx(\Ga)$ in order to obtain $\Klogfxp(\Ga)$ is in one-one 
correspondence with the set of non-finite subsets of $\RSF{\ga}$, and thus
exponential in the size of the formula $\ga$, \emph{provided} that $\ga$ has
already been pre-processed, that is, $\ga$ is a disjunction of pure
$\nb$-formulas.
However, the pre-processing procedure itself, rewriting a modal logic
formula into this normal form, involves (at least) an exponential blow-up.
We conjecture that the two steps of the procedure could be merged into
one single algorithm which would produce an axiomatization of size
exponential in the size of the original formula.
We did not pursue this matter further since for our purposes it suffices to
see that the axiomatization is \emph{finite}, and because we believe that
for clarity of exposition our separation of the various steps in the procedure
is preferrable.
\end{remark}

Theorem~\ref{t:sc} in Section~\ref{s:sc} states the soundness and
completeness of the axiom system $\Klogfxp(\Ga)$ with respect to the
Kripke semantics of $\langfx(\Ga)$, and in the same Section we give an
overview of the proof of this result.


\section{Comparing least fixpoints of systems of equations}
\label{s:soe}

This section is devoted to the proof two rather technical results relating
the existence and nature of the least fixpoints of the formulas and
systems of equations that we discussed in the previous section.
The first proposition substantiates our claim that the semi-simple system
of equations $T_{\ga}$, obtained in step~1 in the procedure, represents the
original formula $\ga$, in the sense that in any modal algebra $A$, the 
(parametrized) least fixpoints of $\ga$ and those of $T_{\ga}$ can be
derived from one another.

\begin{proposition}
\label{p:lfpsoe1}
Let $\ga$ be a modal formula in which the variable $x$ only occurs positively,
let $A$ be a modal algebra, and let $\vec{v} \in A^{P}$ be a sequence of
parameters in $A$.
\begin{enumerate}
\item
The least fixpoint $\mu_{Z}.(T^{A}_{\ga})_{\vec{v}}$ exists iff the least
fixpoint $\mu_{x}.\ga^{A}_{\vec{v}}$ exists.
\item
If existing, these least fixpoints are related as follows.
Writing $\mu_{x}.\ga^{A}_{\vec{v}} = a$ and 
$\mu_{Z}.(T^{A}_{\ga})_{\vec{v}} = \set{ b_{z} \mid z \in Z }$, we have
\begin{align}
  \label{eq:c1}
  a & = b_{z_{\ga}} \,,
  \\ \label{eq:c2}
  b_{z_{\psi}} & = \psi^{A}_{\vec{v}}(a)\,, & \mbox{ for all }\psi \in \RL\,.
\end{align}
\item
If $\mu_{Z}.(T^{A}_{\ga})_{\vec{v}}$ is constructive then so is 
$\mu_{x}.\ga^{A}_{\vec{v}}$.
Conversely, if $\mu_{x}.\ga^{A}_{\vec{v}}$ is constructive, then, provided the
operations in $\ga$ are continuous, $\mu_{Z}.(T^{A}_{\ga})_{\vec{v}}$ is
constructive as well.
\end{enumerate}
\end{proposition}

\begin{proof}
Fix $\ga$, $A$ and $\vec{v}$ as in the statement of the proposition.
In order to simplify notation, we write $\ga$ rather than $\ga^{A}_{\vec{v}}$,
and $T$ rather than $(T^{A}_{\ga})_{\vec{v}}$.

First assume that $\mu_{x}.\ga^{A}$ exists, say $a = \mu_{x}.\ga^{A}$.
It follows from Proposition~\ref{p:fpsoe1} that the vector 
$\{ \psi^{A}(a) \mid \psi \in \RL \}$ is a solution of $T^{A}$.
To see that it is in fact the \emph{least} solution, let $\{ b_{\psi} \mid 
\psi \in \RL \}$ be another solution of $T^{A}$.
Then, again by Proposition~\ref{p:fpsoe1}, $b_{\ga}$ is a solution of the
equation $x = \ga^{A}(x)$, and hence by assumption on $a$, we find $a \leq
b_{\ga}$.
From this, a formula induction shows that $\psi^{A}(a) \leq b_{\psi}$, for 
each $\psi \in \RL$.
This proves the direction ($\Rightarrow$) of part~1, and the equation
(\ref{eq:c2}) of part~2.
The other direction of part~1, and the equation (\ref{eq:c1}) of part~2 have
a similar proof.

For the proof of part~3, we consider the approximating sequences 
$\{ (\ga^{A})^{n}(\bot) \mid n\in\om \}$ and $\{
(T^{A})^{n}(\vec{\bot}) \mid n\in\om \}$.
Abbreviate $c_{n} := (\ga^{A})^{n}(\bot)$ and $t_{n} := \pi_{z_{\ga}}(
(T^{A})^{n}(\vec{\bot}))$.
The main claim in the proof is the following.

\begin{claimfirst}
\label{cl:1}
The sequences $(c_{n})_{n\in\om}$ and $(t_{n})_{n\in\om}$ are mutually cofinal:
\begin{enumerate}
\item For all $n\in\om$ there is an $m\in\om$ such that $t_{n} \leq c_{m}$.
\item For all $n\in\om$ there is an $m\in\om$ such that $c_{n} \leq t_{m}$.
\end{enumerate}
\end{claimfirst}

\begin{pfclaim}
For the first statement of the claim, by induction on $n$ we prove that
\begin{equation}
\label{eq:cl11}
\mbox{for all } n\in\om:\; T^{n}(\vec{\bot}) \leq \set{ \psi(\ga^{n}(\bot)) \mid \psi \in \RL}\, .
\end{equation}
The base case is immediate by the fact that $T^{0}(\vec{\bot}) = \vec{\bot}$.
Inductively, for $\chi \in \RL$ we have
\[\begin{array}{lll}
\pi_{\chi}(T^{n+1}(\vec{\bot})) &=& \widehat{\chi}[T^{n}(\vec{\bot})/\vec{z}]
\\ &\leq& \widehat{\chi}[\vec{\psi}[\ga^{n}(\bot)/x]/\vec{z}] 
\\ &=& \chi(\ga^{n}(\bot))\,.
\end{array}\]
This proves (\ref{eq:cl11}), and so in particular we obtain
\[
\mbox{for all } n\in\om:\; 
t_{n+1} = \pi_{\ga}(T^{n}(\vec{\bot})) \leq \ga(\ga^{n}(\bot)) =
\ga^{n+1}(\bot)\,.
\]
From this the first part of the claim is immediate.

Part~2 of the claim is a little harder to prove.
Given a modal formula $\phi$, let $d(\phi)$ denote the modal \emph{depth}
of $\phi$, and put $k:= d(\ga)$.
Then by induction on $n$ we prove that
\begin{equation}
\label{eq:cl12}
\mbox{for all } n\in\om:\; c_{n} \leq t_{kn}\,.
\end{equation}
Whereas the base case of (\ref{eq:cl12}) is immediate by the fact that $c_{0}
= \bot$, for the inductive case we need a subinduction to prove the
following.
\begin{equation}
\label{eq:cl12a}
\mbox{for all $\chi\in \set{x} \cup \RL$: }
\chi^{A}(c_{n}) \leq \pi_{\chi}(T^{kn+d(\chi)}(\vec{\bot}))\,,
\end{equation}
where we let $\pi_{x}$ denote $\pi_{z_{\ga}}$.

The proof of (\ref{eq:cl12a}) proceeds by induction on the depth of $\chi$.
For the \emph{base step} we must have $\chi = x$.
So in this case we see that $\chi^{A}(c_{n}) = c_{n}$, while
$\pi_{\chi}(T^{kn+0}(\vec{\bot})) = 
\pi_{z_{\ga}}(T^{kn}(\vec{\bot})) = t_{kn}$, where the latter equality
is nothing but the definition of $t_{kn}$.
So in this case, \eqref{eq:cl12a} follows from the main inductive hypothesis.

For the \emph{inductive step}, fix a formula $\chi\in\RL$.
We may write $\chi = \widehat{\chi}(\psi_{1},\ldots,\psi_{n})$,
where each $\psi_{i} \in \RL$ has depth properly smaller than $d(\chi)$,
and $\widehat{\chi} = t_{z_{\chi}}$ is a depth 1 formula such that 
\begin{equation}
\label{eq:cl12b}
\mbox{for all } \vec{a} \in A^{Z},
\widehat{\chi}^{A}(a_{\psi_{1}},\ldots,a_{\psi_{n}}) =
\pi_{\chi}(T(\vec{a}))\,.
\end{equation}
Then we obtain
\begin{align*}
\chi^{A}(c_{n}) 
   & = \widehat{\chi}^{A}
   \big((\psi_{1}^{A}(c_{n}),\ldots,\psi_{n}^{A}(c_{n})\big)
   \tag*{by definition of $\widehat{\chi}$}
\\ & \leq \widehat{\chi}^{A}\big(
   \pi_{\psi_{1}}(T^{kn+d(\psi_{1})}(\vec{\bot})), \ldots, 
   \pi_{\psi_{n}}(T^{kn+d(\psi_{n})}(\vec{\bot}))
   \big)
   \tag*{by the IH}
\\ & \leq \widehat{\chi}^{A}\big(
   \pi_{\psi_{1}}(T^{kn+d(\chi)-1}(\vec{\bot})), \ldots, 
   \pi_{\psi_{n}}(T^{kn+d(\chi)-1}(\vec{\bot}))
   \big)
   \tag*{by monotonicity}
\\ & = \pi_{\chi}\big(T(T^{kn+d(\chi)-1}(\vec{\bot}))
   \big)
   \tag*{by (\ref{eq:cl12b})}
\\ & = \pi_{\chi}(T^{kn+d(\chi)}(\vec{\bot}))\,,
\end{align*}
which proves (\ref{eq:cl12a}).

To obtain the inductive case of (\ref{eq:cl12}) from this, take $\chi := \ga$
in (\ref{eq:cl12a}).
This gives
\[
c_{n+1} = \ga(c_{n}) \leq \pi_{\ga}(T^{kn + d(\ga)}(\vec{\bot}))
\leq \pi_{\ga}(T^{(k+1)n}(\vec{\bot})) = t_{(k+1)n},
\]
as required.
\end{pfclaim}

It easily follows from Claim~\ref{cl:1} that 
\begin{equation}
\label{eq:cntn}
\bv_{n\in\om}c_{n} \mbox{ exists iff }
\bv_{n\in\om}t_{n} \mbox{ exists; }
\mbox{ and if existing, } \bv_{n\in\om}c_{n} = \bv_{n\in\om}t_{n}\,.
\end{equation}

Now suppose that $T$ has a constructive fixpoint $\mu_{Z}.T =
\bv_{n\in\om} T^{n}(\vec{\bot})$.
It follows from part~1 that $\mu_{x}.\ga$ exists and that $\mu_{x}. \ga =
\pi_{\ga}(\mu_{Z}.T)$.
But by the continuity of the projection operation $\pi_{\ga}$ we obtain that
$\pi_{\ga}(\mu_{Z}.T) = \bv_{n\in\om} \pi_{\ga}(T^{n}(\vec{\bot})) =
\bv_{n\in\om} t_{n}$, and so by (\ref{eq:cntn}) we may derive that 
$\mu_{x}.\ga = \bv_{n\in\om} \ga^{n}(\bot)$.
That is, $\ga$ has a constructive fixpoint indeed.
\medskip

Conversely, suppose that $\ga$ has a constructive fixpoint: $\mu_{x}.\ga =
\bv_{n\in\om} \ga^{n}(\bot)$; write $c_{\om} := \mu_{x}.\ga$.
Then by (\ref{eq:c2}), $\mu_{Z}.T = \{ \psi^{A}(c_{\om}) \mid \psi \in \RL \}$.
But if all the operations in $\ga$ are continuous, then each $\psi \in \RL$ is
continuous, implying that
\[
\psi^{A}(c_{\om}) = \bv_{n\in\om} \psi(c_{n})\,.
\]
Then it follows from \eqref{eq:cl12a} and the continuity of the projections
that 
\[
\bv_{n\in\om} \psi(c_{n}) \leq \bv_{m\in\om}\pi_{\psi}(T^{m}(\vec{\bot}))
\leq \pi_{\psi}\Big(\bv_{m\in\om}T^{m}(\vec{\bot})\Big)\,.
\]
Since this applies to all formulas $\psi \in \RL$ we obtain that
\[
T \big(\bv_{n\in\om} T^{n}(\vec{\bot})\big) = \bv_{n\in\om} T^{n}(\vec{\bot})\,.
\]
In other words, $T$ has a constructive fixpoint as well.
\end{proof}

The second proposition in this section relates the least fixpoint of a
semi-simple system of equations to that of its simple simulation.
It justifies the third step in the procedure of defining the axiomatization
which we defined in the previous section.

\begin{proposition}
\label{p:lfpsoe2}
Let $T$ be a semi-simple modal system, let $A$ be a modal algebra, and $\vec{v}
\in A^{P}$ a sequence of parameters in $A$.
\begin{enumerate}
\item
If $A$ is complete, then $\mu_{Z}.T^{A}_{\vec{v}}$ and 
$\mu_{Y}.(T^{+})^{A}_{\vec{v}}$ both exist, and they are related as follows.
Writing
$\mu_{Z}.T^{A}_{\vec{v}} = \{ a_{z} \mid z \in Z \}$
and $\mu_{Y}.(T^{+})^{A}_{\vec{v}} = \{ b_{y} \mid y \in Y \}$, we have:
\[\begin{array}{ll}
a_{z} = b_{\{z\}} & \mbox{for } z\in Z
\\ b_{y_{S}} = \bigwedge_{z\in S} a_{z} & \mbox{for } S \in \Pplus(Z)\,.
\end{array}\]
\item
If $\mu_{Y}.(T^{+})^{A}_{\vec{v}}$ exists and is constructive, then
$\mu_{Z}.T^{A}_{\vec{v}}$ exists, and is constructive as well.
Writing, again,
$\mu_{Z}.T^{A}_{\vec{v}} = \{ a_{z} \mid z \in Z \}$
and $\mu_{Y}.(T^{+})^{A}_{\vec{v}} = \{ b_{y} \mid y \in Y \}$, we have:
\[\begin{array}{ll}
a_{z} = b_{\{z\}} & \mbox{for } z\in Z\,.
\end{array}\]
\end{enumerate}
\end{proposition}

\begin{proof}
Part~1 of the proposition is the main statement of Arnold \& Niwi\'nski
in~\cite[\S 9]{arnoldniwinski}.

Part~2 is a special case of Lemma~\ref{l:x} below.
Too see why we may apply this lemma, take $P := A^{Z}$, $Q:= A^{Y}$, and let
$f$ and $g$ be the maps $T^{A}_{\vec{v}}$ and $(T^{+})^{A}_{\vec{v}}$,
respectively.
Let $\iota: A^{Z} \to A^{Y}$ be as in (\ref{eq:iota}), and let $\pi: A^{Y}
\to A^{Z}$ be given by
\[
\pi(\vec{b})(z) := b_{\{z\}}\,.
\]
Then it is obvious that all maps involved are order preserving, that
$\iota(\bot) = \bot$, and that $\pi(\iota(\vec{a})) = \vec{a}$, for all 
$\vec{a} \in A^{Z}$.
It is straightforward to prove that $\iota$ is continuous, and, finally, 
we already discussed the commutativity of the diagram (\ref{diagram:powerset}).
\end{proof}

We have isolated the following lemma from the proof of the previous
Proposition since it may have some independent interest.

\begin{lemma}
\label{l:x}
Let $P,Q$ be posets with a least element $\bot$ and consider a commuting 
diagram of the form
$$
\xygraph{ []*+{P}="UL" (:[rr]*+{P}="UR"^{f})
  :[d]*+{Q}="DL"^{\iota} 
  (:[rr]*+{Q}="DR"^{g})
  "UR":"DR"^{\iota} 
}
$$
where $f$ and $g$ are order preserving, and $\iota$ is continuous and 
preserves $\bot$.
Moreover, suppose that there exists an order preserving $\pi : Q \rTo P$ such
that $\pi \circ \iota$ is the identity on $P$.  
If $g$ has a constructive least prefixpoint $\mu.g$, then $f$ has also
has a constructive least prefixpoint $\mu.f$ given by the formula
\begin{align*}
  \mu.f & = \pi(\mu.g)\,.
\end{align*}
\end{lemma}

\begin{proof}
We shall prove that, for each ordinal $\alpha$, the following holds:
\begin{equation}
\label{eq:PQ}
\mbox{if $g^{\alpha}(\bot)$ exists, then $f^{\alpha}(\bot)$ exists, and
$\iota(f^{\alpha}(\bot)) = g^{\alpha}(\bot)$.}
\end{equation}
Let us first see how to derive the Lemma from this.
To start with, we may infer from (\ref{eq:PQ}) that for all $\al$ such that
$g^{\alpha}(\bot)$ exists, we have 
$f^{\alpha}(\bot) = \pi(\iota(f^{\alpha}(\bot))) = \pi(g^{\alpha}(\bot))$.
So if $g^{\om+1}(\bot) = g^{\om}(\bot)$, then we immediately obtain that 
$f^{\om+1}(\bot) = \pi(g^{\om+1}(\bot)) = \pi(g^{\om}(\bot)) = f^{\om}(\bot)$.
In other words, if $\mu.g$ is constructive then so is $\mu.f$.

We prove (\ref{eq:PQ}) by ordinal induction on $\al$.
If $\alpha = 0$, then $\iota(f^{0}(\bot)) = g^{0}(\bot)$ amounts to saying
that $\iota$ preserves the least element.
If $\al$ is a successor ordinal $\beta +1$, then the existence of 
$f^{\al}(\bot)$ is not an issue.
The second part of (\ref{eq:PQ}) follows from
\begin{align*}
  \iota(f^{\alpha}(\bot)) = \iota(f(f^{\beta}(\bot))) & =
  g(\iota(f^{\beta}(\bot))) = g(g^{\beta}(\bot)) =
  g^{\alpha}(\bot)\,.
\end{align*}
Here the second identity follows by the commutativity of the diagram, and 
the third identity, by the inductive hypothesis.

If $\alpha$ is a limit ordinal then we will prove first that the approximant
$f^{\alpha}(\bot)$ exists. 
We will actually show that $\pi(g^{\alpha}(\bot)) =
\bigvee_{\beta < \alpha} f^{\beta}(\bot)$, so that
$\pi(g^{\alpha}(\bot)) = f^{\alpha}(\bot)$.
  Observe that, for $\beta < \alpha$, $g^{\beta}(\bot) \leq
  g^{\alpha}(\bot)$ implies $f^{\beta}(\bot) =
  \pi(g^{\beta}(\bot))\leq \pi(g^{\alpha}(\bot))$.
  Also, if $f^{\beta}(\bot) \leq x$ for all $\beta < \alpha$, then
  $g^{\beta}(\bot) = \iota(f^{\beta}(\bot)) \leq \iota(x)$, hence
  $g^{\alpha}(\bot) \leq \iota(x)$ and $\pi(g^{\alpha}(\bot)) \leq
  \pi(\iota(x)) = x$.
We are now ready to argue that $\iota(f^{\alpha}(\bot)) = g^{\alpha}(\bot)$:
  \begin{align*}
    \iota(f^{\alpha}(\bot)) & = \iota(\bigvee_{\beta < \alpha}
    f^{\beta}(\bot)) = \bigvee_{\beta < \alpha} \iota(f^{\beta}(\bot))
    = \bigvee_{\beta < \alpha} g^{\beta}(\bot) = g^{\alpha}(\bot)\,,
  \end{align*}
where we need $\iota$ to be continuous in the second identity.
\end{proof}


\section{Soundness and Completeness}
\label{s:sc}

In this section we state the two main soundness and completeness results of
the paper, and we outline the proofs.  

As mentioned already, our completeness proofs are algebraic in nature,
crucially involving the \emph{Lindenbaum-Tarski algebra} $\Li^{\Slog}(X)$
associated with a system $\Slog$ of axioms and deductive rules, and with a set
$X$ of variables. 
In the next two subsections $\Slog$ will denote $\Klogfx(\Ga)$ and
$\Klogfxp(\Ga)$, respectively, and, if $\Slog$ and $X$ are understood, we
shall write simply $\Li$ in place of $\Li^{\Slog}(X)$.
The definition of $\Li$ is based on the standard construction of an algebra
from the syntax of a logic~\cite{blac:moda01}.
The \emph{elements} of this algebra are equivalence classes of the
formulas/terms that are generated from the set $X$ of variables. 
Here two terms $t,s$ are declared to be equivalent if $\vdash t
\leftrightarrow s$ is derivable in the system $\Slog$. 
The \emph{operations} of our Lindenbaum algebra also have a standard definition.
For example we shall have 
\begin{align*}
  [t] \land^{\Li} [s] & = [t \land s] 
  \intertext{or, for the fixpoint connective $\ff_{\ga}$,}
  \ff_{\ga}^{\Li}([t_{1}],\ldots,[t_{n}] )& =
  [\ff_{\ga}(t_{1},\ldots ,t_{n} )]\,.
\end{align*}
Clearly, for the correctness of the latter definition we use the fact that
the congruence rule
$$
\begin{array}{c}
  \{ \; s_{i} \leftrightarrow t_{i}\;\}_{1 \leq i \leq n}
  \\ \hline
  \ff_{\ga}(s_{1},\ldots,s_{n}) \leftrightarrow
  \ff_{\ga}(t_{1},\ldots,t_{n})
\end{array}
$$
is derivable in $\Klogfx(\Ga)$ --- and a fortiori in $\Klogfxp(\Ga)$ ---
as a straightforward derivation reveals.

Lindenbaum-Tarski algebras are of fundamental importance, both logically and
algebraically.
In logic, they are the algebraic incarnation of the associated derivation
system $\Slog$, in the sense that two formulas $s$ and $t$ are equivalent
with respect to $\Slog$ iff the equation $s = t$ holds in the algebra
$\Li^{\Slog}(X)$ (provided that $X$ contains all the variables occurring
in $s$ and $t$).
Algebraically, they are the free algebras in the class of algebraic models
for the logic.

More specifically, in our setting, both $\Li^{\Klogfx(\Ga)}(X)$ and 
$\Li^{\Klogfxp(\Ga)}(X)$ are modal $\ff$-algebras, and, moreover, the latter
algebra is regular.
Also, in both cases, there is a canonical interpretation of the variables
in $X$ as elements in $\Li$, sending the variable $x$ to the equivalence
class $[x]$ of the term $x$.
Now first let $\Li$ be $\Li^{\Klogfx(\Ga)}(X)$ and observe that whenever
$A$ is a modal $\ff$-algebra and $\vec{v} : X \rTo A$ is a valuation
of the variables in $x$ as elements of $A$, then there exists a unique
modal $\ff$-algebra morphism $f : \Li \rTo A$ such that $f[x] =
\vec{v}(x)$ for all $x \in X$.  
In universal algebraic, or categorical terms, $\Li$ is the \emph{free
$\ff$-algebra over $X$}, and this property, freeness, determines $\Li$ up to
isomorphism of modal $\ff$-algebras.
Next, if we let $\Li$ be $\Li^{\Klogfxp(\Ga)}(X)$, then an analogous
property holds: $\Li$ is the \emph{free regular $\ff$-algebra over $X$}.
\medskip

Returning to the proof sketch, we will underpin our completeness results
algebraically by a representation theorem stating that
\begin{theorem}
  \label{theo:main}
  If $X$ is countable, then $\Li(X)$ embeds in a Kripke $\ff$-algebra.
\end{theorem}
We shall see that such a theorem holds if $\Slog = \Klogfxp(\Ga)$, so
that $\Li(X)$ is the free regular $\ff$-algebra over $X$, or if $\Slog
= \Klogfx(\Ga)$ is the standard Kozen-Park axiomatization and all the
formulas in $\Gamma$ are subject to some syntactic constraints.

\medskip

In both cases, such a result implies completeness as follows.
Let $X$ be the set of variables of a term/formula $t$.  
If the formula $t$ is valid in every Kripke frame, then the equation $t =
\top$ holds in every Kripke \ff-algebra, and thus certainly in the one that
$\Li(X)$ embeds into. 
Consequently, the equation $t = \top$ holds in the Lindenbaum algebra
$\Li(X)$, and by our earlier observation that $\Li$ incarnates the associated
logic, this means that the formula $\top \leftrightarrow t$ is a derivable
theorem of the associated logic.
As usual, this implies that $\vdash t$ is derivable as well, which
establishes the completeness of the logic.

\medskip

In turn, the proof of Theorem \ref{theo:main} is subdivided in many
steps, which we here collect into some main results, to be proved
successively in the next two sections.
\begin{enumerate}
\item First we show that the modal operators $\pos[i]^{\Li}$ of $\Li$ are
  residuated (Corollary~\ref{c:Lires}).
\item
  Then we prove that $\Li$ is constructive (Theorem~\ref{t:constr}).
\item
  Finally, Theorem~\ref{t:4:1} states that every countable modal $\ff$-algebra 
  that has residuated modalities and constructive fixpoint connectives, can be
  embedded in a Kripke $\ff$-algebra.
\end{enumerate}
Since $\Li(X)$ is countable whenever $X$ is countable,
Theorem~\ref{theo:main} follows immediately from this.

\smallskip 

The proof of Theorem \ref{theo:main} will be carried out almost in
parallel for the two systems $\Klogfx(\Ga)$ and $\Klogfxp(\Ga)$. For
the sake of readability, we shall give the details of the proof in the
monomodal setting but discuss also in extent the steps that have to be
taken to generalize the proof to the polymodal setting.

\subsection{Completeness of the Kozen-Park axiomatization}

As we mentioned in the Introduction, in many cases the relatively
simple Kozen axiomatization is already sound and complete with respect
to the Kripke semantics.  This applies to flat modal fixpoint
languages in which each connective $\ff_{\gamma}$ can be defined as
the least fixpoint of a formula $\ga'$ which is \emph{untied} with
respect to $x$.  This notion is closely related to those of the
aconjunctive formulas of Kozen~\cite{kozen} and the disjunctive
formulas of Walukiewicz~\cite{walukiewicz}, but it is fine-tuned to
the fact that we are focussing on the special role of the variable
$x$.

\begin{definition}
  \label{d:untied}
  A modal $\nb$-formula $\ga(x) \in \MT(X)$ is \emph{untied in $x$} if
  it can be obtained from the following grammar:
  \[
  \phi ::= 
  x \mid \top \mid \phi\lor\phi \mid \psi \land \bw_{j\in J} \nab[j]\Phi_{j} 
  \]
Here $\psi$ is a formula in which $x$ does not occur, $J \subseteq I$, and
each $\Phi_{j}$ is a set of $x$-untied formulas.
\end{definition}

\begin{example}
The key point of untied formulas in $x$ is that we restrict the use of
conjunctions to formulas of the form $\psi \land \bw_{j\in J} \nab[j]\Phi_{j}$
where $x$ may not occur in $\psi$, and no two $\nb$ operators in
$\bw_{j\in J} \nab[j]\Phi_{j}$ may be indexed by the same atomic action.
Thus, for instance, the formulas
$\nab[1]\big\{\nab[2]\{p\}\big\} \land \nab[1]\{x\}$
and
$\nab[1]\big\{\nab[2]\{x\}\big\} \land \nab[2]\{x\}$
are untied in $x$,
but the formula 
$\nab[1]\big\{\nab[2]\{x\}\big\} \land \nab[1]\{x\}$
is not.
For a slightly more elaborate example, the formula
\begin{align}
    \label{eq:exampleuntied}
    \phi := & (\,\nab[1]\set{\top,x,\nab[1]\set{\top,x}} \land \nab[2]\nada\,) \vee
    (\,\nab[1]\set{\top,x,\nab[1]\set{\top,x}} \land
    \nab[2]\set{\nab[1]\set{x,\top}} \,)
\end{align}
can be parsed by the above grammar and therefore is untied in $x$.
\end{example}

We can now formulate the first result, returning to its proof at the
end of this subsection.
\begin{theorem}
  \label{t:scs}
  Suppose that each $\ga(x) \in \Ga$ is untied with respect to $x$.
  Then the axiom system $\Klogfx(\Ga)$ is sound and complete with
  respect to the Kripke semantics of $\langfx(\Ga)$.
\end{theorem}

For readers that are not familiar with the cover modalities, we give a
corollary of Theorem~\ref{t:scs} that is phrased in terms of the
classical presentation of modal logic using diamonds and boxes.  We
leave it for the reader to verify that this corollary covers fixpoint
connectives $\ff_{\ga}$ indexed by a formula $\ga$ in which $x$ has
exactly one, positive, occurrence.  This takes care of for instance
the computation tree logic, $\CTL$.

\begin{definition}
  \label{d:harmless}
  A modal formula $\ga(x)$ is \emph{harmless with respect to $x$} if
  it can be obtained from the following grammar:
  \[
  \phi ::= 
  x \mid \top \mid \phi\lor\phi \mid \psi \land \phi \mid
  \dia_{i}\phi \mid \Box_{i}\phi \mid
  \bw_{j\in J}\phi_{j} \,.
  \]
Here $\psi$ is a formula in which $x$ does not occur, $J \subseteq I$,
  and $\bw_{j\in J}\phi_{j}$ is a \emph{harmless conjunction}.  This
  means that for each $j \in J$, the conjunct $\phi_{j}$ is either of
  the form $\Box_{j}\chi$, or itself a conjunction of the form
  $\bw_{\ell\in L}\pos[j]\chi_{\ell}$ (with $\chi$ and each
  $\chi_{\ell}$ being harmless in $x$).
\end{definition}

\begin{example}
\label{ex:harmless}
The formula $\dia_{1}(x \land \dia_{2} x)$ is not harmless, and neither is
$\dia_{1} x \land \Box_{1}\dia_{2}x$. The formula $\dia_{1} x \land
\dia_{1}\dia_{1}x \land \Box_{2}\dia_{1}x$ is, on the other hand,
harmless, and this also applies to
$\dia_{1} x \land \dia_{1}\dia_{1}x \land \Box_{1}\dia_{1}p$.
\end{example}

\begin{corollary}
\label{c:scsc}
Let $\Ga$ be a set of modal formulas each of which is harmless with respect
to $x$.
Then the axiom system $\Klogfx(\Ga)$ is sound and complete with respect to the
Kripke semantics of $\langfx(\Ga)$.
\end{corollary}

\begin{proof}
A straightforward induction shows that every $\ga(x)$ which is harmless 
with respect to $x$, is equivalent to a $\nb$-formula that is untied in $x$.
(For instance, the harmless formula $\dia_{1} x \land \Box_{1}\dia_{2}x$ of
Example~\ref{ex:harmless} is equivalent to the untied formula 
\eqref{eq:exampleuntied}.)
Then the Corollary is immediate by Theorem~\ref{t:scs}.
\end{proof}

\begin{proofof}{Theorem ~\ref{t:scs}}
  The axiomatization $\Klogfx(\Gamma)$ certainly is sound. To argue
  about completeness, we need Theorem~\ref{theo:main} for $\Li$ the
  Lindenbaum algebra associated with $\Klogfx(\Gamma)$.  We proceed
  along the path sketched above and, to this goal, the key observation
  is that if $\ga$ is untied in $x$, then for any vector of parameters
  $\vec{v}$, the term function $\ga^{\Li}_{\vec{v}}$ on the
  Lindenbaum-Tarski algebra $\Li = \Li^{\Klogfx(\Ga)}(X)$ is a
  finitary $\O$-adjoint. 
This implies that the least fixpoint $\ga^{\Li}_{\vec{v}}$ is constructive,
see Theorem~\ref{t:constr} for more details of this argument.
\end{proofof}

\subsection{The general case}

We leave it as an open problem whether, in the general case, the
system $\Klogfx(\Ga)$ is complete.  However, for its extension
$\Klogfxp(\Ga)$ we have the following uniform soundness and
completeness result.

\begin{theorem}
  \label{t:sc}
  The axiom system $\Klogfxp(\Ga)$ is sound and complete with respect
  to the Kripke semantics of $\langfx(\Ga)$.
\end{theorem}

In the sequel we shall use the phrase ``free
regular $\ff$-algebra'' as a synonym of the Lindenbaum algebra, and
$\Li$, $\Li(X)$ shall be short notation for $\Li^{\Klogfxp(\Gamma)}(X)$.
\medskip
\begin{proofof}{Theorem ~\ref{t:sc}}
  As we mentioned already in the previous section, the \emph{soundness} of
  $\Klogfxp(\Ga)$ follows from the main result of Arnold \& Niwi\'nski 
  in~\cite[\S 9]{arnoldniwinski}, here mentioned as Proposition~\ref{p:lfpsoe2}.
  For, it is an immediate consequence of this result that all Kripke
  $\ff$-algebras are regular.  But from this and the fact that the
  Lindenbaum-Tarski algebra is the free regular $\ff$-algebra, the
  soundness of $\Klogfxp(\Ga)$ follows by a standard algebraic logic
  argument.

To argue for completeness, we need Theorem~\ref{theo:main} for $\Li$ the
Lindenbaum algebra associated with $\Klogfxp(\Gamma)$. We proceed
  again along the path sketched above but this time the path is less direct.

To argue that the least fixpoint of $\gamma^{\Li}_{\vec{v}}$ is constructive,
we first observe that the least fixpoint of $(T^{+}_{\gamma})^{\Li}_{\vec{v}}$
-- which by regularity exists -- is constructive, since $(T^{+}_{\gamma})$ is
a simple system and its interpretation in $\Li$ is a finitary $\O$-adjoint.
Then, the property of constructiveness of the respective fixpoints can be
transferred from $(T^{+}_{\gamma})^{\Li}_{\vec{v}}$ to
$(T_{\gamma})^{\Li}_{\vec{v}}$ and from
  $(T_{\gamma})^{\Li}_{\vec{v}}$ to $\gamma^{\Li}_{\vec{v}}$, using
  the results of Section~\ref{s:soe}. A detailed account of this
  process will be given in the proof of Theorem~\ref{t:constr}.
\end{proofof}


\section{Properties of the Lindenbaum Algebras}
\label{s:3}
\label{sec:constructive}
\label{s:constructive}

\newcommand{\A}{\Alg{A}}
\newcommand{\AP}{\Alg{A}^{\Pi}}

The goal of this section is to prove that the Lindenbaum algebra
$\Li^{\Slog}$, where $\Slog$ is one of the axiom systems
$\Klogfx(\Gamma)$ and $\Klogfx(\Gamma)$, is constructive,
cf.~Definition \ref{def:constructive}.  We shall obtain this result by
subsequently analyzing properties of this algebra.  

\subsection{Rigidness}

\noindent
We start with showing that $\Li$ is \emph{rigid} with respect to $X$.

\begin{definition}
Let $\Alg{A}$ be a modal algebra generated by a set $X$.
$\Alg{A}$ is \emph{rigid} with respect to $X$ if
\begin{equation}
\label{eq:whitman}
\bigwedge G \;\land\; \nb Y = \bot 
\timplies 
\bigwedge G = \bot \tor \exists y \in Y \tst y = \bot,
\end{equation}
where $G$
and $Y$ are finite, possibly empty, sets of elements of $\Alg{A}$, with
$G \sse \{ x, \neg x \mid x \in X \}$.
\end{definition}

\begin{remark}
In a polymodal setting we say that $\Alg{A}$ is \emph{rigid} with respect
to $X$ if
  \begin{align*}
    \bigwedge G \;\land\; \bigwedge_{i \in I}\nab[i] Y_{i} = \bot
    \timplies 
    \bigwedge G & = \bot \tor \exists i \in I \tand y \in Y_{i} \tst y =
    \bot\,.
  \end{align*}
\end{remark}
\begin{remark}
  To gather some intuitions about this property, we first prove
  rigidness of the free \emph{modal} algebra $\mathcal{M}(X)$
  generated by a set $X$ of variables.  Reformulating the property in
  terms of formulas, and reasoning by contraposition, it suffices to
  show that whenever $\La$ is a consistent set of $X$-literals, and
  $\Phi = \{ \phi_{1}, \ldots, \phi_{n} \}$ is a set of consistent
  modal formulas, then the formula $\bw\La \land \nb\Phi$ is
  consistent as well.

  So let $\La$ and $\Phi$ be as indicated.  Then by completeness there
  is a pointed Kripke model $(\bbM_{i},r_{i})$ for each formula
  $\phi_{i}$.  Now create a new model $\bbM$ as follows.  Take the
  disjoint union of the models $\bbM_{1},\ldots,\bbM_{n}$, and add one
  single new point $r$.  Let $\{r_{1},\ldots,r_{n}\}$ be the successor
  set of $r$, and define a valuation for $r$ so that the propositional
  formula $\bw\La$ is true at $r$.  Clearly then $\bbM,r \forces
  \bw\La \land \nb\Phi$, witnessing the consistency of the formula
  $\bw\La \land \nb\Phi$.

  Second, for readers that are familiar with the \emph{duality theory}
  of modal algebras~\cite{vene:alge06}, the notion of rigidness has a
  very natural formulation in terms of the dual relational space
  $A_{*}$ of $A$.  Let $A$ be a modal algebra generated by some set
  $X$.  Then $A$ is rigid with respect to $X$ iff for every finite set
  $G \sse \{ x,\neg x \mid x \in X \}$ such that $\bw G > \bot$, and
  every finite set $U = \{ u_{1}, \ldots, u_{n} \}$ of ultrafilters in
  $A_{*}$, there is an ultrafilter $u \supseteq G$ which has $U$ as
  its collection of successors.
\end{remark}

\begin{theorem}
  \label{theo:mainsec}
  Let $\Li$ denote either the free modal $\ff$-algebra or the free regular
  modal $\ff$-algebra. Then $\Li$ is rigid with respect to $X$.
\end{theorem}

The proof of the Theorem depends on the following construction.  

\begin{definition}
  \label{d:adduf}
  Let $A$ be some modal algebra, and let $\Pi = \set{ \pi_{\ell}: A
    \rTo 2 \mid 0 < \ell \leq n }$ be a finite (possibly empty) set of
  Boolean algebra homomorphisms.  We define the operation $\dia^{\Pi}:
  A \rTo 2$ by putting
  \begin{align}
    \dia^{\Pi}(a) & := \bigvee \set{ \pi(a) \mid \pi \in \Pi }.
    \label{eq:diapi}
  \end{align}
  We define the operation $\dia^{\AP}: A \times 2 \rTo A \times 2$ as
  follows:
  \begin{align}
    \label{eq:diaap}
    \dia^{\AP}(a,d) & := (\dia^{A}(a), \dia^{\Pi}(a))\,,
  \end{align}
  and let $\AP$ be the algebra obtained by expanding the Boolean algebra
  $\Alg{A}\times 2$ with this operation.
\end{definition}

For future reference we define the cover operation associated with
$\dia^{\Pi}$ as the map $\nb^{\Pi}: \Pom A \rTo 2$ given by
\begin{align}
  \label{eq:nbPi}
  \nb^{\Pi}\al & := \Box^{\Pi}\bv\al \land \bw\dia^{\Pi}\al\,,
\end{align}
where of course $\Box^{\Pi}x = \neg \pos^{\Pi} \neg x$.  

\begin{remark}
In a polymodal setting the construction has to be parameterized by a
collection of the form $\set{\Pi_{i} \mid i \in I }$.
Then $\dia^{\Pi_{i}}$ and $\pos[i]^{\AP}$ are defined from $\Pi_{i}$ as in
the equations \eqref{eq:diapi} and \eqref{eq:diaap}, respectively.
\end{remark}

\begin{remark}
Again, a dual perspective on this construction may be illuminating.
Recall that Boolean homomorphisms may be identified with ultrafilters.
In a nutshell (and again, presupposing familiarity with the duality theory of
modal algebras), we obtain the dual structure of $A^{\Pi}$ by \emph{adding}
an ultrafilter $u$ to the dual structure $A_{*}$ of $A$, making the set $\Pi$
of Boolean homomorphisms/ultrafilters its successor set.
\end{remark}

It is not difficult to verify that the operation $\dia^{\AP}$ is additive,
so that $\AP$ is a modal algebra. 
But in fact, as we will see in the Proposition below, the construction
preserves many other properties as well.

\begin{proposition}
\label{p:rigid1}
Let $\A$ be a modal algebra, and let  $\Pi = \{ \pi_{i}: A \rTo 2 \mid 0 < i
\leq n \}$ be a finite set of Boolean algebra homomorphisms.
\begin{enumerate}
\item 
If $A$ is a modal $\ff$-algebra for some fixpoint connective $\ff_{\ga}$, then
so is $\AP$.
\item
Let $T$ be a semi-simple modal system, and let $\vec{v} \in A^{P}$ be some 
parameter for $T$.
If $T^{A}_{\vec{v}}$ has a least solution on $A$, then so does 
$T^{\AP}_{(\vec{v},\vec{w})}$, for each parameter $\vec{w} \in 2^{P}$.
\item If $\A$ is regular with respect to some semi-simple modal system $T$,
then so is $\AP$.
\end{enumerate}
\end{proposition}

\begin{proof}
  Since part~1 of the proposition is a direct consequence of part~2
  and Proposition~\ref{p:lfpsoe1}, we start with proving part~2.  Let
  $T = \langle Z, \{ t_{z} \mid z \in Z \}\rangle$ be a semi-simple
  system of equations.  Since the carrier of $\AP$ is the set $A
  \times 2$, we may see $T^{\AP}: (\AP)^{Z} \times (\AP)^{P} \rTo
  (\AP)^{Z}$ as a map
  \[
  T^{\AP}:
  (A^{Z}\times A^{P}) \times (2^{Z}\times 2^{P}) \rTo (A^{Z}\times 2^{Z})\,.
  \]
  Let $\pi_{A}$ and $\pi_{2}$ denote the projections of $\AP$ onto $A$
  and $2$, respectively.
  
  Given the definition of the modal operator of $\AP$, the first
  coordinate $\pi_{A} \circ T^{\AP}$ of the map $T^{\AP}$ is identical
  to the map $T^{A} \circ \pi_{A}$.  Furthermore, since $T$ is
  semi-simple, in each term $t_{z}$ the unguarded variables are all
  from $P$, while the guarded variables are all from $Z$, and each
  occurrence of these is in the scope of exactly one modality.  
As a consequence, the second coordinate 
of $T^{\AP}$ is the compose of
  \[
  (A^{Z}\times 2^{Z}) \times (A^{P}\times 2^{P})
  \stackrel{\pi}{\rTo} A^{Z} \times 2^{P} 
  \stackrel{\wt{T_{2}}}{\rTo} 2^{Z}.
  \]
  Here $\wt{T_{2}}$ is best understood by observing that its terms are
  obtained from those of $T$ by replacing every occurrence of the
  symbol $\nb$ with the formal symbol $\nb^{\Pi}$.

  Summarizing, we may write $T^{\AP} = \langle T^{A} \circ \pi_{A},
  \wt{T_{2}} \circ \pi \rangle$.  It follows by Beki\v{c}' property
  that, for each $\vec{v}\in A^{P}$ and $\vec{w} \in 2^{P}$, the least
  fixpoint of $T^{\AP}_{(\vec{v},\vec{w})}$ exists, and can be
  written as
  \begin{equation}
    \label{eq:lfpAP}
    \mu_{Z}.T^{\AP}_{(\vec{v},\vec{w})} \,=\,
    \langle
    \mu_{Z}.T^{A}_{\vec{v}}, \wt{T_{2}} (\mu_{Z}.T^{A}_{\vec{v}}, \vec{w}) \rangle.
  \end{equation}
  
  Part~3 also follows from part~2, but it needs more work.
  We first prove that the following diagram commutes, for every $\vec{w} \in
  2^{P}$:
  \begin{align}
    \label{diag:commwidetilde}
    &
    \xygraph{
      []*+{A^{Z}}="U" 
      (:[rrr]*{\,\,\,2^{Z}}="RU"^{(\wt{T_{2}})_{\vec{w}}}) 
      :[dd]*+{A^{Y}}^{\iota^{A}} 
      :[rrr]*+{2^{Y}}="RD"^{(\wt{T^{+}_{2}})_{\vec{w}}}
      "RU":"RD"^{\iota^{2}}
    }
  \end{align}
  Recall that in Section~\ref{s:ax} we showed the diagram
  (\ref{diagram:powerset}) to commute because of
  Proposition~\ref{p:simterm}(2).  A careful analysis of that
  proposition reveals that the only property needed for its proof is
  that the diamond $\dia$ underlying the operation $\nb$ (in the sense
  that $\nb \al = \Box\bv\al \land \bw\dia\al$ with $\Box a =
  \neg\dia\neg a$) preserves finite joins.  Now the operation
  $\dia^{\Pi}$ underlying the operation $\nb^{\Pi}$ of $\wt{T_{2}}$
  also preserves finite joins, and so we prove that the diagram
  (\ref{diag:commwidetilde}) commutes in exactly the same manner.

  Now we establish the regularity of $\AP$ as follows.  First, it
  follows from part~2 of this proposition that for each $\vec{v}\in
  A^{P}$ and $\vec{w} \in 2^{P}$, the least fixpoint
  $\mu_{Y}.(T^{+})^{\AP}_{(\vec{v},\vec{w})}$ exists.  Moreover, we
  may calculate
  \begin{align*}
    \mu_{Y}.(T^{+})^{\AP}_{(\vec{v},\vec{w})}
    & = \mylangle
    \mu_{Y}.(T^{+})^{A}_{\vec{v}}
    \mycomma
    \wt{T^{+}_{2}}(
    \mu_{Y}.(T^{+})^{A}_{\vec{v}},\vec{w})
    \myrangle
    \tag*{by (\ref{eq:lfpAP}),}
    \\
    & = \mylangle   
    \iota^{A}(\mu_{Z}.T^{A}_{\vec{v}})
    \mycomma
    \wt{T^+_{2}}(
    \iota^{A}(\mu_{Z}.T^{A}_{\vec{v}}),\vec{w})
    \myrangle 
    \tag*{since $A$ is regular,}
    \\
    & = \mylangle   
    \iota^{A}(\mu_{Z}.T^{A}_{\vec{v}})
    \mycomma
    \iota^{2}(\wt{T_{2}}(
    \mu_{Z}.T^{A}_{\vec{v}}, \vec{w}))
    \myrangle 
    \tag*{since diagram \eqref{diag:commwidetilde} commutes,}
    \\
    & = \iota^{\AP}(
    \mylangle  \mu_{Z}.T^{A}_{\vec{v}}
    \mycomma 
    \wt{T_{2}}(
    \mu_{Z}.T^{A}_{\vec{v}}, \vec{w}) 
    \myrangle
    )
    \tag*{since $\iota^{\AP} = \iota^{A} \times \iota^{2}$,}
    \\
    & = \iota^{\AP}(
    \mu_{Z}.T^{A}(\vec{v},\vec{w}))
    \tag*{again, by (\ref{eq:lfpAP}).}
  \end{align*}
  This finishes the proof of the third and final part of the
  proposition.
\end{proof}

We can now prove the rigidness of $\Li$, on the basis of
Proposition~\ref{p:rigid1} and the fact that $\Li$ is the \emph{free}
$\ff$-algebra over $X$. Moreover, part 3 of Proposition~\ref{p:rigid1}
ensures that the same proof works if $\Li$ is the \emph{free regular}
$\ff$-algebra over $X$.

\begin{proofof}{Theorem~\ref{theo:mainsec}}
  Suppose for contradiction that $\Li$ is not rigid with respect to
  $X$.  Then there is a finite set $\La$ of $X$-literals, and a finite
  subset $\al \sse_{\om} A$ such that $\bigwedge\La > \bot$ and $b >
  \bot$ for all $b \in \al$, while $\bigwedge\La \land \nb^{\Li} \al =
  \bot$.

  By the prime filter theorem, we may find a set $\Pi = \{ \pi_{b}:
  \Li \rTo 2 \mid b \in \al \}$ of Boolean homomorphisms such that
  $\pi_{b}(b) = \top$ for all $b \in \al$.
Now consider the algebra $\Li^{\Pi}$, and let $f: X \to \Li^{\Pi}$ be some
map satisfying
\begin{equation}
\label{eq:deff}
  f(x) = \left\{\begin{array}{ll}
      (x,\top) & \mbox{if $x \in \La$}
      \\
      (x,\bot) & \mbox{if $\neg x \in \La$.}
    \end{array}\right.
\end{equation}
  Clearly, such a map exists by the consistency of $\La$, and since
  $\Li$ is the free (regular) $\ff$-algebra generated by $X$, $f$ can be
  extended to a modal $\ff$-homomorphism $\wt{f}$ from $\Li$ to
  $\Li^{\Pi}$.  Then it follows from our assumption that
  $\wt{f}(\bigwedge\La \land \nb^{\Li} \al) = \wt{f}(\bot^{\Li}) =
  \bot^{\Li^{\Pi}}$.

  On the other hand, we claim that 
  \begin{equation}
    \label{eq:adduf}
    \wt{f}(\bw\La \land \nb^{\Li}\al) = (\bot,\top),
  \end{equation}
  which provides us with the desired contradiction.  For the proof of
  (\ref{eq:adduf}), using the fact that $\wt{f}$ is a homomorphism, we
  find
  \[
  \wt{f}(\bw\La \land \nb^{\Li}\al) 
  = \wt{f}(\bw\La) \land \wt{f}(\nb^{\Li}\al).
  \]
From the assumption \eqref{eq:deff} on $f$, and the fact that $\wt{f}$ is an 
extension of $f$, it follows that $\wt{f}(a)
  = (a,\top)$ for all $a$ in $\La$, so that $\wt{f}(\bw\La) = \bw \{
  \wt{f}(a) \mid a \in \La \} = (\bw\La,\top)$, while
  $\wt{f}(\nb^{\Li}\al) = (\nb^{\Li}\al, \nb^{\Pi}\al)$, where
  $\nb^{\Pi}$ is the cover modality associated with $\dia^{\Pi}$, see
  (\ref{eq:nbPi}).  
The point of the construction of $\Li^{\Pi}$ is that
\begin{equation}
\label{eq:point}
\nb^{\Pi}\al = \top,
\end{equation}
as we shall prove now.
The relation \eqref{eq:point} trivially holds if $\alpha$ is empty, since then
$\dia^{\Pi} x = \bot$ for all $x \in A$ and so $\nb^{\Pi}\al =
\Box^{\Pi}\bot = \neg \dia^{\Pi} \top = \top$. 
So let us now assume that $\alpha$ is not empty.  
Then we compute 
\begin{align*}
\Box^{\Pi}\bv\al 
   &= \neg\dia^{\Pi}\big(\bw\{ \neg b \mid b \in \al \}\big)
\\ & = \neg \bv_{a \in \al} \pi_{a} \big(\bw\{ \neg b \mid b \in \al \}\big)
   \tag*{by (\ref{eq:diapi})}
\\ & \geq \neg \bv_{a \in \al} \pi_{a} ( \neg a)
   \tag*{($\pi_{a}$ is monotone)}
\\ & \geq \neg \bv_{a \in \al} \neg\pi_{a} (a)
   \tag*{($\pi_{a}$ is a homomorphism)}
\\ & = \neg \bv_{a \in \al} \neg\top
   \tag*{(by assumption on $\pi_{a}$)}
\\ &
    = \top
  \end{align*}
  and
  \begin{align*}
    \bv\dia^{\Pi}\al &= \bv_{b \in \al} \dia^{\Pi} b 
    = \bv_{b\in\al} \bv_{a\in\al} \pi_{a}(b) 
    \geq \bv_{b\in\al} \pi_{b}(b) 
    = \bv_{b\in\al} \top 
    = \top
  \end{align*}
  so that we find
  \[
  \nb^{\Pi}\al = \Box^{\Pi}\bv\al \land \bv\dia^{\Pi}\al = \top,
  \]
  which proves (\ref{eq:point}).  Continuing our computation of
  $\wt{f}(\bw\La \land \nb^{\Li}\al)$, we now have that
  \[
  \wt{f}(\bw\La \land \nb^{\Li}\al) = 
  (\bw\La,\top) \land (\nb^{\Li}\al,\top) =
  (\bw\La \land \nb^{\Li}\al,\top \land \top) =
  (\bot,\top)\,.
  \]
  
  \noindent
  This finishes the proof of \eqref{eq:adduf}, and thus, of the Theorem.
\end{proofof}
\begin{remark}  
In a polymodal setting, by the same sort of computations, we shall have
  $$
  \wt{f}(\bw\La \land \nbb^{\Li}\vec{\al}) = 
  (\bw\La,\top) \land \bw_{i \in I}(\nab[i]^{\Li}\al_{i},\top) =
  (\bw\La \land \nb^{\Li}\al,\top \land \bw_{i \in I} \top) =
  (\bot,\top)\,.
  $$
  Thus, in presence of many modalities, a contradiction with
  the regularity of $\Li$ is obtained in a similar way.
\end{remark}

\subsection{Finitary $\O$-adjoints}

We now turn to the notion of a finitary $\O$-adjoint and to its
generalization, that of a finitary family of $\O$-adjoints, see
Definition \ref{def:finitaryOadj}.
The use of these notions lies in an earlier result by the first
author~\cite{sant:comp08}, which roughly states that fixpoints of
finitary $\O$-adjoints, if existing, are constructive.  In order to
apply this result we aim to show that simple systems of equations on
the Lindenbaum algebra give rise to finitary $\O$-adjoints.
To reach this goal we only need $\L$ to be rigid with respect to $X$ and
to be generated by $X$. Therefore the next results apply both to the
Lindenbaum algebra $\Li^{\Klogfx(\Gamma)}$ and to the Lindenbaum
algebra $\Li^{\Klogfxp(\Gamma)}$.

Our first observation is that the cover modality $\nb^{\Li}$ on the
Lindenbaum algebra is itself a finitary $\O$-adjoint.  
In order to turn this into a meaningful mathematical statement, we need to
endow the domain $\Pom(\Li)$ of the operation $\nb^{\Li}$ with a
quasi-order, see Remark~\ref{rem:quasiorder}.  Thus, let us define the
relation $\rl{\leq}$ on $\Pom(\Li)$ by saying that $\al\rl{\leq} \be$
iff for all $a\in\al$ there is a $b\in\be$ such that $a\leq b$, and
for all $b\in\be$ there is an $a \in \al$ such that $a \leq b$.  It is
not hard to see that $\rl{\leq}$ is a quasi-order on $\Pom(\Li)$.

\begin{theorem}
\label{t:nbOadj}
Let $\Li$ denote either the free modal $\ff$-algebra or the free regular modal
$\ff$-algebra.
Then each cover modality ${\nab_{i}}^{\Li}: \Pom(\Li) \rTo \Li$ is an $\O$-adjoint.
\end{theorem}

\begin{proof}
  Given an element $d \in \Li$, we need to define a finite set
  $G_{\nb}(d) \in \Pom\Pom(\Li)$ such that for all $\al \in
  \Pom(\Li)$, we have
  \begin{equation}
    \label{eq:nbOadj}
    \nb\al \leq d \mbox{ iff } \al\rl{\leq}\be \mbox{ for some } \be\in G_{\nb}(d)\,.
  \end{equation}
  
  First we confine our attention to the so-called \emph{weakly
    irreducible elements} of $\Li$, that is, the ones of the form
  \begin{equation}
    \label{eq:wie}
    \bv \Pi \lor \dia b \lor \bv_{c\in C}\Box c\,,
  \end{equation}
  where $\Pi$ is some set of $X$-literals, $b$ is an element of $\Li$,
  and $C$ is a finite set of elements of $\Li$.
  
  For a weakly irreducible element $d = \bv \Pi \lor \dia b \lor
  \bv_{c\in C}\Box c$ we let
  \begin{align}
    G_{\nb}(d) & := G^{\Pi}(d) \cup G^{\dia}(d) \cup G^{\Box}(d)\,,
    \label{eq:Gofwirr}
    \intertext{where}
    \label{eq:defgipi}
    G^{\Pi}(d) &:= 
    \left\{\begin{array}{ll}
        \big\{ \{\top\}, \nada \big\} & \mbox{if } \bv\Pi = \top,
        \\ \nada & \mbox{otherwise},
      \end{array}\right.
    \\[3mm]
    \nonumber
    G^{\dia}(d) &:= 
    \big\{ \{ b, \top \} \big\}\,,
    \\[1mm]
    \nonumber
    G^{\Box}(d) &:= \bigcup_{c \in C} \set{\{ b \vee c \} , \nada}\,.
  \end{align}
  The correctness of this definition follows from the following Claim.
  
  \begin{claimfirst}
    Let $d = \bv \Pi \lor \dia b \lor \bv_{c\in C}\Box c$ be weakly
    irreducible.  Then the following are equivalent, for any $\al \in
    \Pom(\Li)$:
    \begin{enumerate}
    \item $\nb\al \leq d$;
    \item \begin{enumerate}
      \item $\bv\Pi = \top$, or
      \item $a \leq b$ for some $a \in \al$, or
      \item $\bv\al \leq b \vee c$ for some $c \in C$;
      \end{enumerate}
    \item $\al \rl{\leq}\be$, for some $\be \in G_{\nb}(d)$.
    \end{enumerate}
  \end{claimfirst}
  
\begin{pfclaim}
(1 $\Rightarrow$ 2) Reasoning by contraposition, we assume that (2) does not
hold.  
Then (a$'$) the set $\La := \{ \neg\pi \mid \pi\in\Pi \}$ of literals is
consistent, (b$'$) $\neg b \land a > \bot$ for every $a \in \al$, and (c$'$)
$\neg b \land \neg c \land \bv\al > \bot$ for every $c \in C$.  
Now consider the element
\[
e := \bw\La \land \nb\big( \{ \neg b \land  a \mid a \in\al \} \cup
    \{ \neg b \land \neg c \land \bv\al \mid c \in C \} \big).
\]
It is immediate that $e \leq \bw\La$, and easy to verify that $e \leq
\bw_{c\in C} \dia\neg c$.
In addition, considering that 
\begin{eqnarray*}
e &\leq& \Box\bv \big( \{ \neg b \land  a \mid a \in\al \} \cup
    \{ \neg b \land \neg c \land \bv\al \mid c \in C \} \big)
\\ &=& \Box \big( \neg b \land \bv 
(\{ a \mid a \in\al \} \cup
    \{ \neg c \land \bv\al \mid c \in C \}) \big)
\end{eqnarray*}
we have $e \leq \Box\neg b$.
Combining these observations, we find that $e \leq \neg d$.
But it is also easily seen that $e \leq \nb\al$.
On the other hand, we may apply the rigidness of $\Li$ to derive from
(a$'$)--(c$'$) that $e >\bot$.
From this it follows that $\nb\al\not\leq d$; that is, (1) fails, as required.

    \noindent
    (2 $\Rightarrow$ 1) In each of the cases (2a)--(2c) it is obvious that
    $\nb\al \leq d$.
    
    \noindent
    (2 $\Rightarrow$ 3)
    Suppose that (2) holds, and distinguish cases.
    (a) If $\bv\Pi = \top$ then both $\{\top\}$ and $\nada$ belong to
    $G_{\nb}(d)$.  Then $\al \rl{\leq} \{ \top\}$ if $\al\neq\nada$,
    and $\al\rl{\leq}\nada$ if $\al = \nada$, so there is always some
    $\be\in G_{\nb}(d)$ with $\al\rl{\leq} \be$.
    (b) If $a\leq b$ for some $a\in\al$, then it is easy to see that
    $\al \rl{\leq} \{b,\top\}$, and this suffices to prove (3) since
    in this case $\{ b,\top \}$ belongs to $G_{\nb}(d)$.
    (c) If $\bv\al \leq b \vee c$, with $c \in C$, then $\al\rl{\leq}
    \{b \vee c \}$ if $\alpha
    \neq \nada$ and $\alpha \rl{\leq} \nada$ if  $\alpha
    = \nada$.
    In both cases we have proved (3), since both $\nada$ and $\{b \vee c\}$
    belong to $G_{\nb}(d)$.
    
    \noindent
    (3 $\Rightarrow$ 2) Assume that $\al\rl{\leq}\be$, with $\be\in
    G_{\nb}(d)$, and again distinguish cases.  If $\be \in
    G^{\Pi}(d)$, then in particular $G^{\Pi}(d)$ is nonempty; this can
    only be the case if $\bv\Pi = \top$, so (2a) holds.  If $\be \in
    G^{\dia}(d)$, then $\be = \{ b,\top \}$, so from $\al\rl{\leq}\be$
    it follows that there is some $a\in\al$ such that $a\leq b$, so
    (2b) holds.  Finally, if $\be \in G^{\Box}(d)$, then $C$ is not
    empty. If $\beta = \nada$, then $\alpha = \nada$. Let $c \in C$ be
    arbitray, then $\bv \alpha = \bot \leq b \vee c$. If $\be =\{b
    \vee c\}$ for some $c \in C$, then from $\al\rl{\leq}\be$ we may
    deduce that $a\leq b \vee c$ for all $a\in\al$.  This implies
    $\bv\al \leq b \vee c$. In both cases (2c) holds.
  \end{pfclaim}
  
  Finally, let $d$ be an arbitrary element of $\Li$.  It is not hard
  to show that $d$ can be written as a finite meet $d = \bw_{\ell =1,\ldots ,n}
  d_{\ell}$ of weakly irreducible elements.  Thus, in order to define
  $G_{\nb}(d)$ for such a meet, it is enough to define
  $G_{\nb}(\top)$ and $G_{\nb}(d_{1} \land d_{2})$ assuming that we
  have already defined $G_{\nb}(d_{1})$ and $G_{\nb}(d_{2})$. We let
  \begin{align}
    \label{eqs:Gofmeet}
    G_{\nb}(\top) & = \set{\set{\top},\emptyset}\,, \\
    \nonumber
    G_{\nb}(d_{1}\land d_{2}) & =
    \set{ \set{b_{1} \land b_{2}\mid (b_{1},b_{2}) \in Z} \mid
      \exists \beta_{i} \in G_{\nb}(d_{i}),i = 1,2, \tand Z \in \beta_{1}
      \bowtie \beta_{2} }\,.
  \end{align}
We leave it to the reader to verify that, with the above definition,
$G_{\nb}(\top)$ satisfies (\ref{eq:nbOadj}).
For $G_{\nb}(d_{1}\land d_{2})$ we argue as follows. If $\nb \alpha
  \leq d_{1} \land d_{2}$ then, for $i = 1,2$, $\nb \alpha \leq d_{i}$
  and $\alpha \rl{\leq} \beta_{i}$ for some $\beta_{i} \in
  G_{\nb}(d_{i})$.  
Define $Z$ by putting $(b_{1},b_{2}) \in Z$ iff there exists $a \in \alpha$
such that $a \leq b_{1}$ and $a \leq b_{2}$.
Then $Z \in \beta_{1} \bowtie \beta_{2}$ and $\alpha \rl{\leq} \set{b_{1}
\land b_{2} \mid (b_{1},b_{2}) \in Z}$. Conversely, if for $i=1,2$, some
  $\beta_{i} \in G_{\nb}(d_{i})$ and some $Z \in \beta_{1} \bowtie
  \beta_{2}$, the relation $\alpha \rl{\leq} \set{b_{1} \land b_{2}
    \mid (b_{1},b_{2}) \in Z}$ holds, then $\alpha \rl{\leq}
  \beta_{i}$, so that $\nab \alpha \leq d_{i}$, $i = 1,2$, and $\nab
  \alpha \leq d_{1}\land d_{2}$.
\end{proof}
\begin{remark}
  In a polymodal setting the vectorial nabla $\nbb = \bw_{i\in I}
  \nab[i]$ is an $\O$-adjoint on the Lindenbaum algebra
  $\Li$. Recalling that $\nbb^{\Li} : \Pom(\Li)^{I} \rTo \Li$, then we
  need to define $G_{\nbb^{\Li}}(d)$ as a finite set of vectors (of
  finite subsets of $\Li)$, that is, $G_{\nbb^{\Li}}(d) \subseteq_{\om}
  \Pom(\Li)^{I}$. 
To this aim, we proceed as before: we first define
  it on weakly irreducible elements and then we extend its definition
  to meets of weakly irreducible elements.  Now, in a polymodal setting, $d$ is
  weakly irreducible if it can be written as
  \begin{align*}
    d & = \bv \Pi \vee \bv_{i \in I} (\,\pos[i]b_{i} \vee \bigvee
    \nec[i] C_{i}\,)\,.
  \end{align*}
  For $d$ weakly irreducible, we let
  \begin{align*}
    G_{\nbb^{\Li}}(d) & = \vec{G}^{\Pi}(d) \cup \bigcup_{i \in I}
    \vec{G}_{i}(d)
    \intertext{where}
    \vec{\beta} \in \vec{G}^{\Pi}(d) & \tiff
    \vec{\beta}_{i} \in G^{\Pi}(d) \text{ for all } i \in I\,,  \\
    \vec{\beta} \in \vec{G}_{i}(d) & \tiff
    \vec{\beta}_{i} \in \set{\set{b_{i},\top}} \cup \set{\set{b_{i}
        \vee c} \mid
      c \in C_{i}}
    \tand \vec{\beta}_{k} \in
    \set{ \set{\top},\nada} \text{ for } k \neq i\,,  %
  \end{align*}
  where $G^{\Pi}(d)$ is defined as in equation \eqref{eq:defgipi}.

To see that this is a correct definition, it suffices to observe that $\nbb
\vec{\beta} \leq d$ if $\vec{\beta} \in G_{\nbb^{\Li}}(d)$, and that, 
conversely, $\nbb^{\Li} \vec{\alpha} \leq d$ implies the existence of some
$\vec{\beta} \in G_{\nbb^{\Li}}(d)$ such that $\alpha_{i} \rl{\leq} \beta_{i}$
for all $i \in I$.
The first of these two observations is straightforward; the second follows
from an analog to Claim~1 in the proof of Theorem~\ref{t:nbOadj} stating
that by the rigidness of $\Li$, $\nbb^{\Li} \vec{\alpha} \leq d$ implies
one of the following three cases: (1) either $\bv \Pi = \top$, or (2) there
exists $i \in I$ and $a \in \vec{\alpha}_{i}$ such that $a \leq b_{i}$, or
(3) there exists $i \in I$ and $c \in C_{i}$ such that $\bigvee 
\vec{\alpha}_{i} \leq b_{i} \vee c$.

\smallskip

  To extend the definition of $G_{\nbb^{\Li}}$ to all elements of
  $\Li$, we let
  \begin{align*}
    \vec{\beta} \in G_{\nbb^{\Li}}(\top) & \tiff \vec{\beta}_{i} \in
    \set{\set{\top},\nada} \text{ forall }
    i \in I\,, \\
    \vec{\beta} \in G_{\nbb^{\Li}}(d_{1} \land d_{2}) & \tiff \exists
    \vec{\beta}^{j} \in G_{\nbb^{\Li}}(d_{j}), j = 1,2, \\
    & \tand Z_{i}
    \in \vec{\beta}^{1}_{i} \bowtie \vec{\beta}^{2}_{i} \tst
    \vec{\beta}_{i} = \set{b_{1} \land b_{2} \mid (b_{1},b_{2}) \in
      Z_{i}}\,.
  \end{align*}
We leave it for the reader to verify the correcteness of this definition
along the ideas given for formulas \eqref{eqs:Gofmeet}.
\end{remark}

As an immediate corollary of Theorem~\ref{t:nbOadj}, we obtain the
following.

\begin{corollary}
  \label{c:Lires}
The Lindenbaum algebra $\Li$ is residuated, that is, each operation
$\dia_{i}^{\Li} : \Li \rTo \Li$ is a left adjoint.
\end{corollary}

\begin{proof}
Recall that $\dia x = \nb \{ x,\top\}$ and observe that the correspondence
$\set{\cdot,\top}: \Li \to \Pom(\Li)$, sending $x \in \Li$ to $\set{x,\top}
\in \Pom(\Li)$ is an $\O$-adjoint: 
We can define
  \begin{align*}
    G_{\set{\cdot,\top}}
    & =
    \begin{cases}
      \set{\bigwedge \alpha} \,,& \top \in \alpha \,,\\
      \emptyset\,,& \text{otherwise}\,,
    \end{cases}
\end{align*}
leaving it for the reader that this definition is indeed correct.
  As $\O$-adjoints compose, it follows from Theorem~\ref{t:nbOadj}
  that $\dia^{\Li}$ is an $\O$-adjoint.  But then it must be a left
  adjoint since it preserves finite joins, see \cite[Proposition
  6.3]{sant:comp08}.
\end{proof}

\begin{remark}
In passing we note that the same results apply to the free \emph{modal}
algebra, which can be identified with the Lindenbaum-Tarski algebra
of the basic (poly-)modal logic $\Klog$.
In particular, simplified versions of the proofs given here will show that
the coalgebraic modality of the free modal algebra is an $\O$-adjoint.
\end{remark}

In order to prove the main result of this section, viz.,
Proposition~\ref{p:shortSoE} dealing with constructiveness of simple systems
of equations, we need to adapt the definition of the cover modality so that
it has as its domain a product set of the form $A^{Z}$. Formally, for a
finite set of variables $Z$, we introduce the operation $\nb^{A}_{Z} :
A^{Z} \rTo A$, defined by the formula
\begin{align*}
  \nb_{Z}(\vec{v}) & = \bigwedge_{z \in Z} \pos \vec{v}_{z} \land \nec
  \bigvee_{z \in Z}\vec{v}_{z}\,.
\end{align*}
If $Y \subseteq Z$, then we shall write $\nb_{Y}^{A} : A^{Z} \rTo A$
for the compose $\nb_{Y}^{A} \circ \pi_{Y}$, where $\pi_{Y} : A^{Z}
\rTo A^{Y}$ denotes the obvious projection.

It is not difficult to see that $\nb^{A}_{Z} = \nb^{A} \circ
S^{A}_{Z}$, where $S_{Z}^{A} : A^{Z} \rTo \Pom(A)$ transforms a vector
into a finite subset, $S_{Z}^{A}(\vec{v}) = \set{\vec{v}_{z} \mid z
  \in Z}$.
Now, $S_{Z}^{A}$ is an $\O$-adjoint for every modal algebra $A$, since
we can define
\begin{align}
  \label{eq:GofS}
  G_{S_{Z}^{A}}(\beta) & = \set{ \vec{v}^{R} \mid
    R \in Z \bowtie \beta
  }\,,
  \;\;\text{with}\;\;
  \vec{v}^{R}_{z}
   = \bigwedge \set{b \in \beta \mid z R b }\,.
\end{align}

The first part of the next Lemma is an immediate consequence of our
previous observations. The second part of the Lemma will be needed
when arguing about constructiveness of a simple system of equations.

\begin{lemma}
  \label{lem:formofG}
  For every pair $(Z,Y)$ with $Z$ a finite set of variables $Z$ and $Y
  \subseteq Z$, the following holds:
  \begin{enumerate}\item 
    The vectorial cover modality $\nb^{\Li}_{Y} : \Li^{Z} \rTo \Li$ is
    an $\O$-adjoint on the Lindenbaum algebra $\Li$.
  \item Let $d = \bigwedge_{\ell = 1,\ldots ,n} d_{\ell}$, where each
    $d_{\ell}$ is a weakly irreducible element of the form $\bigvee
    \Lambda_{\ell} \vee \pos b_{\ell} \vee \bigvee \nec C_{\ell}$.  If
    $\vec{v} \in G_{\nb^{\Li}_{Y}}(d)$ and $z \in Z$, then
    $\vec{v}_{z}$ is a conjunction of elements from the set
    $\bigcup_{\ell =1,\ldots ,n} \set{b_{\ell}} \cup \set{b_{\ell}\vee
      c \mid c \in C_{\ell}}$.
  \end{enumerate}
\end{lemma}

\begin{proof}
The first part of the Lemma is an immediate consequence of the facts that
$\pi^{\Li}_{Y}$, $S^{\Li}_{Y}$, and $\nb^{\Li}$ are all $\O$-adjoints,
that $\O$-adjoints compose, and that
$\nb^{\Li}_{Y} = 
\nb^{\Li} \circ S^{\Li}_{Z} \circ \pi^{\Li}_{Y}$:
\[
\Li^{Z} \stackrel{\pi^{\Li}_{Y}}{\rTo}
\Li^{Y} \stackrel{S^{\Li}_{Z}}{\rTo}
\Pom(\Li) \stackrel{\nb^{\Li}}{\rTo} \Li\,.
\]

  For the second part of the Lemma we argue as follows.  Let $D$ be
  the set $\bigcup_{\ell =1,\ldots ,n} \set{b_{\ell}} \cup
  \set{b_{\ell}\vee c \mid c \in C_{\ell}}$.
From the equations \eqref{eq:Gofwirr} and \eqref{eqs:Gofmeet} we prove,
  by induction on $n$, that if $a \in \alpha \in
  G_{\nb}^{\Li}(\bigwedge d_{\ell})$, then $a$ is a (possibly empty)
  conjunction of elements from $D$. Then we use the formula that
  witnesses that $\O$-adjoints compose, $G_{g\circ f}(d) = \bigcup_{c
    \in G_{f}(d)} G_{g}(c)$ and the expressions for $G_{S^{\Li}_{Y}}$
  and $G_{\pi^{\Li}_{Y}}$.  From equation \eqref{eq:GofS} it is
  immediately seen that if $\vec{v} \in G_{S_{Y}^{\Li}\circ
    \nb^{\Li}}(d)$ and $y \in Y$, then $\vec{v}_{y}$ is a conjunction
  of elements from $D$.
We leave it for the reader to determine an expression
  for $G_{\pi^{\Li}_{Y}}$ and to conclude that $\vec{v}_{z}$ is a
  conjunction of elements from $D$ if $\vec{v} \in
  G_{\nb^{\Li}_{Y}}(d)$ and $z \in Z$.
\end{proof}

On the basis of the results obtained until now, we can use Proposition 6.3
of~\cite{sant:comp08} to prove that, if $T = \langle Z, \set{t_{z}\mid z
  \in Z}\rangle$ is a \emph{simple} system of equations, then
$T^{\Li}_{\vec{v}}: \Li^{Z} \rTo \Li^{Z}$ is an $\O$-adjoint, for each
parameter $\vec{v}$. However, our real goal is to argue
that $T^{\Li}_{\vec{v}}$ is a \emph{finitary} $\O$-adjoint and hence,
by Proposition \ref{p:foadjconstructive}, that the least fixpoint
$\mu_{Z}.T^{\Li}_{\vec{v}}$ is constructive. To this goal, we shift
the focus of our discussion from $\O$-adjoints to families of
$\O$-adjoints.

\begin{definition}
  A modal algebra $A$ is said to be \emph{$\nb$-finitary} if any
  family $\mathcal{F}$ of the form
  \begin{equation}
    \label{eq:shffoa}
    \mathcal{F} = \set{ k_{\ell} \land \nb^{A}_{Y_{\ell}}: A^{Z} \rTo A \mid
    \ell = 1,\ldots ,n }
  \end{equation}
  is a finitary family of $\O$-adjoints -- where $Z$ is a finite set
  of variables and for each $\ell =1,\ldots ,n$ $Y_{\ell} \subseteq Z$
  and $k_{\ell} \in A$.
\end{definition}

\begin{proposition}
  \label{p:Lishort}
  The Lindenbaum algebra $\Li$ is $\nb$-finitary.
\end{proposition}
\begin{proof}
  Let us define the Fischer-Ladner closure
  $\FL(\phi)$ of a formula $\phi$ as the least set of formulas
  satisfying  the following equations:
  \begin{align*}
    FL(p) & = \set{p} \\
    FL(\neg \phi) & = \set{\neg \phi} \cup FL(\phi) \\
    FL(\phi_{1} \land \phi_{2}) & = \set{\phi_{1} \land \phi_{2}} \cup
    FL(\phi_{1}) \cup FL(\phi_{2}) \\
    FL(\pos \phi) & = \set{\pos \phi} \cup FL(\phi) \\
    FL(\sharp_{\gamma}(\vec{\phi}))
    & = \set{\sharp_{\gamma}(\vec{\phi})} 
    \cup FL(\gamma(\sharp_{\gamma}(\vec{\phi}),\vec{\phi}))\,.
  \end{align*}
  It is a standard argument to prove that $FL(\phi)$ is a finite set.

  Next,  consider a family $\mathcal{F}$ as in equation
  \eqref{eq:shffoa}.
  We shall first argue that the family of $\O$-adjoints
  $$
  \mathcal{F}' = \set{\nb^{\Li}_{Y_{\ell}} : \Li^{Z} \rTo \Li \mid \ell =
    1,\ldots ,n} \cup \set{k_{\ell} \land \cdot : \Li \rTo \Li \mid \ell =
    1,\ldots ,n}
  $$ 
  is finitary. To this goal, we fix an arbitrary formula $\phi_{0}$
  and need to construct a finite set $V$ such that $[\phi_{0}] \in V$
  and $V$ is $\mathcal{F}'$-closed.  We begin by fixing formulas
  $\phi_{\ell}$, $\ell =1,\ldots ,n$, such that $[\phi_{\ell}] =
  k_{\ell}$. Next we let $V \subseteq \Li$ be the Boolean algebra
  generated by the set $\bigcup_{\ell=0,\ldots ,n} \set{[\psi] \mid
    \psi \in FL(\phi_{\ell}) }$. Clearly $V$ is finite and contains
  $[\phi_{0}]$. In order to show that $V$ is $\mathcal{F}'$-closed, we
  observe first that $V$ is generated by the modal equivalence
  classes, i.e. equivalence classes $[\psi]$, where $\psi \in
  \bigcup_{\ell = 0,\ldots ,n} FL(\phi_{\ell})$ is such that $\psi = p$ is a
  propositional variable or $\psi = \pos \psi'$ for some $\psi' \in
  \bigcup_{\ell = 0,\ldots ,n} FL(\phi_{\ell})$.  Hence, if $d \in V$, then
  $d$ is a conjunction of disjunctions of modal equivalence classes
  and their negations. Therefore $d$ is a conjunction of weakly
  irreducible elements of the form $\bigvee \Lambda \vee \pos b \vee
  \bigvee_{c \in C} \nec c$ with $\set{b} \cup C \subseteq V$.

  We can now argue that $V$ is $\mathcal{F}'$-closed. If $d \in V$,
  then write $d$ as a conjunction of weakly irreducible elements
  $d_{j}$ of the form $\bigvee \Lambda_{j} \vee \pos b_{j} \vee
  \bigvee_{c \in C_{j}} \nec c$ with $\set{b_{j}} \cup C_{j} \subseteq
  V$. Then, by Lemma \ref{lem:formofG}, if $z \in Z$ and $\vec{v} \in
  G_{\nb_{Y}^{\Li}}(d)$, then $\vec{v}_{z} \in V$, since $\vec{v}_{z}$
  is a conjunction of elements that belong to $\bigcup_{j} \set{b_{j}}
  \cup \set{b_{j} \vee v \mid c \in C_{j}}$, so that $\vec{v}_{z} \in
  V$. This shows that $V$ is $\nb_{Y_{\ell}}^{\Li}$-closed.  Similarly,
  since the map $(k_{\ell} \land \cdot)$ is left adjoint to the map
  $(\neg k_{\ell} \vee \cdot)$, $G_{k_{\ell} \land \cdot}(d) = \set{\neg
    k_{\ell} \vee d} = \set{\neg [\phi_{\ell}] \vee d} \subseteq V$ provided
  $d \in V$. This shows that $V$ is also $(k_{\ell} \land \cdot)$-closed,
  and therefore we have established that $\mathcal{F}'$ is a finitary
  family.

  Finally, since finitary families are closed under composition and a
  sub-family of a finitary family is a finitary family, see
  Proposition \ref{prop:propertiesoadjoints}, we may deduce that
  $\mathcal{F}$ is itself a finitary family of $\O$-adjoints.
\end{proof}

\begin{proposition}
  \label{p:shortSoE}
  Let $T = \langle Z,\set{t_{z} | z \in Z}\rangle$ be a simple system of
  equations, let $A$ be a $\nb$-finitary modal algebra, and let
  $\vec{v}$ be a set of parameters for $T$.  Then
  $\mu_{Z}.T^{A}_{\vec{v}}$, if existing, is constructive.
\end{proposition}

\begin{proof}
  Let $T$, $A$, and $\vec{v}$ be as stated, and recall that each
  $(t_{z})_{\vec{v}}^{A}$ is of the form $\bigvee_{\ell \in L} k_{\ell}
  \land \nb_{Y_{\ell}}^{A}$. Since families of finitary $\O$-adjoints can
  be closed under joins, it follows from the assumptions that the family
  \[
  \{ (t^{A}_{z})_{\vec{v}}: A^{Z} \rTo A \mid z \in Z \}
  \]
  is a family of finitary $\O$-adjoints.  Hence, by Proposition
  \ref{prop:propertiesoadjoints}, $T^{A}_{\vec{v}}$ is itself a finitary $\O$-adjoint, and
  hence its least fixpoint, if existing, is constructive by
  Proposition \ref{p:foadjconstructive}.
\end{proof}

As a specific example of Proposition~\ref{p:shortSoE}, we see that on
a regular $\ff$-algebra, the modal system $T^{+}_{\ga}$ is
constructive.  Together with the results in Section~\ref{s:soe}, this
is the key to prove constructiveness of the least fixpoint
$\ff_{\ga}$ itself.

\subsection{Constructiveness of $\Li$}

We have now gathered sufficient material to prove the main result of this
section.

\begin{theorem}
  \label{t:constr}
  The Lindenbaum algebra $\Li$ of the system $\Klogfxp(\Gamma)$ is
  constructive.  If every $\gamma \in \Gamma$ is equivalent to an
  untied formula, then the Lindenbaum algebra $\Li$ of the simpler system
  $\Klogfx(\Gamma)$ is constructive.
\end{theorem}

\begin{proof}
  For the first part of the statement we argue as follows.
  We have seen in Section \ref{s:sc} that $\Li$ is the free regular
  modal $\ff$-algebra.  In particular $\Li$ is regular and
  $(T^{+}_{\ga})_{\vec{v}}$ has a least fixpoint
  $\mu_{Z}.(T^{+}_{\ga})_{\vec{v}}$ for each parameter $\vec{v} \in
  \Li^{P}$.  Since $T^{+}_{\ga}$ is a \emph{simple} system of
  equations, it follows from Proposition 
  \ref{p:shortSoE} that each of these least fixpoints
  $\mu_{Z}.(T^{+}_{\ga})_{\vec{v}}$ is constructive.  But then it
  follows by successive applications of the
  Propositions~\ref{p:lfpsoe2} and~\ref{p:lfpsoe1} that all
  parametrized least fixpoints on $\Li$ of $T_{\ga}$ and $\ga$,
  respectively, are constructive as well.

  The second part is even simpler: $\Li$ is, in this case, the free
  modal $\ff$-algebra. Being rigid, the operations that can be
  constructed using substitution starting from $\nb^{\Li}$, constants,
  conjunctions with constants, and disjunctions, are finitary
  $\O$-adjoints on $\Li$. 
  If $\gamma \in \Gamma$ -- so that $\gamma$ is untied -- then
  $\gamma^{\Li}(x,\vec{p})$ is among these operations. Thus,
  $\gamma^{\Li}_{\vec{p}}(x)$ is a finitary $\O$-adjoint and its least
  fixpoint is constructive.
\end{proof}



\section{A representation theorem}
\label{s:4}
\label{sec:embedding}
\label{s:embedding}

\newcommand{\eps}{\epsilon}
\newcommand{\diap}{\blacklozenge}
\newcommand{\cp}{{\neg}}
\newcommand{\rn}{r_{\!{N}}}

\newcommand{\D}{\widetilde{L}}
\newcommand{\Du}{\Delta_{\uparrow}}
\newcommand{\Dd}{\Delta_{\downarrow}}
\newcommand{\Ddm}[1]{\Delta_{\downarrow,-#1}}

The aim of this section is to prove that every countable modal
$\ff$-algebra $A$ in which each diamond modality is residuated, and
each fixpoint connective is constructiuve, can be embedded in a Kripke
$\ff$-algebra (Theorem~\ref{t:4:1} below).  Our proof method consists
of building a representation for $A$ via a step-by-step approximation
process and can be seen as a version of more general game-based
methods for building structures in model theory
(see~\cite{hodg:mode93,hodk:rela00} for an overview).  It has a long
history in modal and algebraic logic,
see~\cite{lynd:repr50,madd:some82,burg:axio82} for some early
references.

\begin{theorem}
  \label{t:4:1}
  Let $A$ be a countable modal \ff-algebra.  Assume that each
  $\ff_{\ga}$ is constructive on $A$, and that each $\mop{i}^{A}$ is
  residuated.  Then $A$ can be embedded in a Kripke \ff-algebra.
\end{theorem}

Fix an algebra $A$ as in Theorem~\ref{t:4:1}.
\newcommand{\reftotheo}{t:4:1} For simplicity we restrict attention to
a language with a single diamond $\dia$, and a single fixpoint
connective $\ff$.  We let $\gamma(x,\vec{p})$ denote the associated
formula of $\ff$, where $\vec{p} = (p_{1},\ldots,p_{n})$.
We will say that $a \in A$ is \emph{nonzero} if $a \neq \bot$.

\noindent
The main lemma in the proof of Theorem~\ref{\reftotheo} is the
following.

\begin{lemma}
\label{l:4:1}
For each nonzero 
$a \in A$ there is a Kripke frame $S_{a}$ and a modal
\ff-homo\-morphism $\rho_{a}: A \to S_{a}^{\ff}$ such that
$\rho_{a}(a) > \bot$.
\end{lemma}

The key notion involved in the step-by-step approximation process
leading up to Lemma~\ref{l:4:1} is that of a \emph{network}.  Let
$\om^{*}$ denote the set of finite sequences of natural numbers.  We
denote concatenation of such sequences by juxtaposition, and write
$\eps$ for the empty sequence.  If $t = sk$ for some $k\in\om$ we say
that $s$ is the parent of $t$ and write either $s = t^{-}$ or $s \lhd
t$.  A \emph{tree} is a subset $T$ of $\om^{*}$ which is both downward
and leftward closed; that is, if $t \neq \eps$ belongs to $T$, then so
does $t^{-}$, and if $sm \in T$ then $sk \in T$ for all $k<m$.
Obviously, a tree $T$, together with the relation $\lhd$, forms a Kripke
frame; this frame will also be denoted as $T$, and its complex \ff-algebra,
as $T^{\ff}$.

An \emph{$A$-network} is a pair $N = \struc{T,L}$ such that $T$ is a
tree and $L: T \to \power(A)$ is some labelling.  Such a network $N$
induces a map $\rn: A \to \power(T)$, given by
\begin{equation}
\label{eq:rep}
\rn(a) := \set{ t \in T \mid a \in L(t) }\,.
\end{equation}
The aim of the proof will be to construct, for an arbitrary nonzero $a
\in A$, a network $N = \struc{T,L}$, with $a \in L(\eps)$, and such
that $\rn$ is a modal \ff-homomorphism from $A$ to $T^{\ff}$.  We
need some definitions.

A network $N = \struc{T,L}$ is called \emph{locally coherent} if $\bw
X > \bot$, whenever $X$ is a finite subset of $L(t)$ for some $t \in
T$; \emph{modally coherent} if $\bw X \land \dia\bw Y > \bot$, for all
$s,t \in T$ such that $s \lhd t$ and all finite subsets $X$ and $Y$ of
respectively $L(s)$ and $L(t)$; and \emph{coherent} if it satisfies
both coherence conditions.  \ $N$ is \emph{prophetic} if for every $s
\in T$, and for every $\dia a \in L(s)$, there is a \emph{witness} $t
\in T$ such that $s \lhd t$ and $a \in L(t)$; \emph{decisive} if
either $a \in L(t)$ or $\cp a \in L(t)$, for every $t \in T$ and $a
\in A$; 
and
\emph{$\ff$-constructive} if, for every $t \in T$, and every sequence
$\vec{a}$ in $A$ such that $\ff\vec{a} \in L(t)$, there is a natural
number $n$ such that $(\ga^{A}_{\vec{a}})^{n}(\bot) \in L(t)$.  A
network is \emph{perfect} if it has all of the above properties.

\begin{lemma}
  \label{l:4:2}
  If $N$ is a perfect $A$-network, then $\rn : A \rTo T^{\ff}$ is a modal
  \ff-homomorphism from the modal $\ff$-algebra $A$ to the complex
  algebra $T^{\ff}$ of the Kripke model $\struc{T,\lhd}$.
\end{lemma}
Clearly, we shall have that $\rn(a) \neq \nada$ for all $a \in A$ for
which there is a $t \in T$ with $a \in L(t)$.
\begin{proof}
  Let $N = \struc{T,L}$ be a perfect network.  It is fairly easy to
  derive from local coherence and decisiveness that each $L(t)$ is an
  \emph{ultrafilter} of (the Boolean reduct of) $A$.  From this it is
  immediate that $\rn$ is a Boolean homomorphism.

  In order to prove that $\rn$ is a modal homomorphism, we need to show
  that
\begin{equation}
  \label{eq:rmh}
  \rn(\dia a) = \set{ t \in T \mid t \lhd s \mbox{ for some } s \in \rn(a) }\,,
\end{equation}
for all $a \in A$.
The inclusion $\sse$ holds because $N$ is prophetic.
For the opposite inclusion, assume that $t \lhd s$ and $a \in L(s)$.
Suppose for contradiction that $t \not\in \rn(\dia a)$, so that $\dia a
\not\in L(t)$.  
Then by decisiveness, $\cp\dia a \in  L(t)$.
This gives the desired contradiction with the assumed modal coherence of 
$N$, so that indeed we may conclude that \eqref{eq:rmh} holds.

  From this it follows that, for all sequences $\vec{a} \in A^{n}$,
  and all modal formula $\phi$:
  \begin{equation}
    \label{eq:rfmahm}
    \phi^{T^{\ff}}(\rn(\vec{a})) = \rn(\phi^{A}(\vec{a}))\,,
  \end{equation}
  where for a vector $\vec{a} = (a_{1},\ldots,a_{n})$ $\rn(\vec{a})$
  denotes -- here and in the sequel -- the vector $(\rn(a_{1}),\ldots,
  \rn(a_{n}))$.

  In particular, for $\phi=\ga$, \eqref{eq:rfmahm} implies that for
  all $\vec{b}$:
  \[
  \rn(\ff^{A}\vec{b}) = \rn(\ga^{A}(\ff^{A}\vec{b},\vec{b})) =
  \ga^{T^{\ff}}(\rn(\ff^{A}\vec{b}),\rn(\vec{b}))\,.
  \]
  In other words, $\rn(\ff^{A}\vec{b})$ is a \emph{fixpoint} of the map
  $\ga^{T^{\ff}}_{\rn(\vec{b})}$.  But we can also prove that
  $\rn(\ff^{A}\vec{b})$ is the $\om$-approximation of
  $\ff^{T^{\ff}}(\rn(\vec{b}))$.  To see why this is so, we start from
  the definition of $\rn(\ff^{A}\vec{b})$:
  \begin{equation}
    \rn(\ff^{A}\vec{b}) = \set{ t \in T \mid \ff\vec{b} \in L(t) }\,.
  \end{equation}
  Since $L(t)$ is an ultrafilter and the network $T$ is
  \ff-constructive, $\ff\vec{b} \in L(t)$ if and only if, for some
  $n$, $(\ga^{A}_{\vec{b}})^{n}(\bot) \in L(t)$, and hence
  \begin{equation}
    \label{eq:rnn}
    \rn(\ff^{A}\vec{b}) = \bigcup_{n<\om}\set{ t \in T \mid 
    \left(\ga^{A}_{\vec{b}}\right)^{n}(\bot) \in L(t) }\,.
  \end{equation}
Recall that, by  definition of $\rn$,
$\left(\ga^{A}_{\vec{b}}\right)^{n}(\bot) \in L(t)$ if and only if
$t \in \rn(\left(\ga^{A}_{\vec{b}}\right)^{n}(\bot))$. 
Moreover, a straightforward inductive proof, on the basis of
\eqref{eq:rfmahm}, will show that 
$$
\rn(\left(\ga^{A}_{\vec{b}}\right)^{n}(\bot))
= \left(\ga^{T^{\ff}}_{\rn(\vec{b})}\right)^{n}(\bot)\,.
$$
  Hence equation \eqref{eq:rnn} becomes
  \[
  \rn(\ff^{A}\vec{b}) = \bigcup_{n<\om} \left(
    \ga^{T^{\ff}}_{\rn(\vec{b})} \right)^{n} (\bot^{T^{\ff}})\,.
  \]
  
  But if $\rn(\ff^{A}\vec{b})$ is both a fixpoint of the map
  $\ga^{T^{\ff}}_{\rn(\vec{b})}$ \emph{and} an ordinal approximation
  of $\ff^{A}(\rn(\vec{b}))$, then it must be the \emph{least} fixpoint
  of the map $\ga^{T^{\ff}}_{\rn(\vec{b})}$, or, equivalently,
  \[
  \rn(\ff^{A}\vec{b}) = \ff^{T^{\ff}}(\rn(\vec{b})).
  \]
  Having shown that $\rn$ is also a homomorphism with respect to
  $\ff$, we have completed the proof of the Lemma.
\end{proof}
From the previous Lemma it follows that, in order to prove
Lemma~\ref{l:4:1}, it suffices to construct a perfect network with $a
\in L(\eps)$ for an arbitrary nonzero $a \in A$.  Our construction
will be carried out in a step-by-step process, where at each stage we
are dealing with a finite \emph{approximation} of the final network.
Since these approximations are not perfect themselves, they will
suffer from certain \emph{defects}.  We will only be interested in
those defects that can be \emph{repaired} in the sense that the
network can be extended to a bigger version that is lacking the
defect.

Formally we define a \emph{defect} of a network $N = \struc{T,L}$ to
be an object $d$ of one of the following three kinds:
\begin{enumerate}
\item
$d = (t,a,\cp)$, with $t \in T$ and $a \in A$ such that neither $a$ nor $\cp 
a$ belongs to $L(t)$,
\item $d = (t,a,\dia)$, with $t \in T$ and $a \in A$ such that $\dia a
  \in L(t)$, but there is no witness $s$ such that $t \lhd s$ and $a
  \in L(s)$,
\item $d = (t,\vec{a},\ff)$, with $t \in T$ and $\vec{a} \in A^{n}$
  such that $\ff\vec{a} \in L(t)$, but there is no $n \in \om$ such
  that $(\ga^{A}_{\vec{a}})^{n}(\bot) \in L(t)$.
\end{enumerate}
These three types of defects witness a network's failure to be decisive,
prophetic, and \ff-constructive, respectively.

In our proof we will construct a perfect network as a limit of
coherent networks, one by one repairing the defects of the
approximants.  In order to guarantee the coherence of these
approximants in the long run, we need them to satisfy a stronger,
\emph{global} version of coherency.  To define this notion we extend
the \emph{local} labelling function $L$ of the network to a global
one, $\D$.  This global labelling gathers all relevant information
concerning the network at one single node.  Since $N$ is finite, it is
straightforward to define such a global labelling map for the
\emph{root} $\eps$ of the tree: if we let
\[
\Dd(t) := \bw L(t) \land \bw_{t\lhd s}\dia\Dd(s)\,,
\]
then the set $\Dd(\eps)$ on its own collects all relevant information
from the full network.  The \emph{residuatedness} of the modality
$\dia$ allows us to access the global information on the network at
\emph{each} of its nodes, not just at the root.  The resulting
labelling $\D: T \rTo A$ will considerably simplify the process of
repairing defects.

Turning to the technical details, for the definition of $\D$ we use the
\emph{conjugate} of $\dia$, which can be defined as the unique map $\diap:
A \rTo A$ satisfying
\begin{equation}
  \label{eq:cj}
  a \land \dia b > \bot \mbox{ iff } \diap a \land b > \bot\,,
\end{equation}
for all $a,b \in A$.  This map exists by the fact that $\dia$ is
residuated; in fact, it is the Boolean dual of the residual (or right
adjoint) of $\dia$.  Using this operation $\diap$, we can define the
global labelling $\D$ as follows:
\begin{eqnarray*}
  \D(t)  &:=& \Dd(t) \land \Du(t),
  \\ \Dd(t) &:=& \bw L(t) \land \bw_{t\lhd s}\dia\Dd(s),
  \\ \Du(t) &:=& \left\{\begin{array}{ll}
      \top & \mbox{if } t = \eps,
      \\ \diap (\Du(t^{-}) \land \Ddm{t}(t^{-})) & \mbox{otherwise},
    \end{array}\right.
  \\ \Ddm{u}(t) &:=& \bw L(t) \land \bw_{t \rhd s,\,s \neq u}\dia\Dd(s)\,.
\end{eqnarray*}
The idea behind this definition is straightforward: for $\D(t)$,
we start by collecting the local information $\bw L(t)$ and then move on
to $t$'s neighbors, both its predecessor (with $\Du(t)$) and its
successors (with $\Dd(s)$).  The role of $\Ddm{u}$ is to ensure
termination of the procedure, avoiding a loop between $\Du(t)$ and
$\Dd(u)$ when $t \lhd u$.

Alternatively, we can understand the formula for $\D(t)$ as follows.
Given $t \in T$, we consider the unoriented tree $T'$ which is obtained by
forgetting the orientation of the edges of the form $u \lhd v$. 
Using a basic result in graph theory, we obtain a unique new orientation
$\rightarrow$ on $T'$ by taking $t$ as a new root.
Observe that $u \rightarrow v$ implies that either $u \lhd v$ or $v \lhd u$.
Then the formula for $\D(u)$ can be defined inductively on the basis of the
new orientation, analogous to the definition of $\Dd(u)$, with the proviso 
that the conjunct contributed by a $\rightarrow$-successor $v$ is modalized
by $\dia$ if $u\lhd v$, and by $\diap$ if $v \lhd u$.
More precisely:
\[
\D(u) = \bw L(u) 
    \land \bw \big\{ \dia \D(v) \mid u \rightarrow v \;\&\; u\lhd v \big\}
    \land \bw \big\{ \diap\D(v) \mid u \rightarrow v \;\&\; v\lhd u \big\}\,.
\]

One of the key observations in the proof is the following claim.
\begin{lemma}
\label{l:4:3}
Let $N$ be a finite network.
Then $\D(s) > \bot$ iff $\D(t) > \bot$, for any $s,t \in N$.
\end{lemma}
\begin{proof}
  It clearly suffices to prove the following special case:
  \begin{equation}
    \label{eq:4c}
    \D(t) > \bot \mbox{ iff } \D(t^{-}) > \bot,
  \end{equation}
  for an arbitrary $t \neq \eps$.  But it is straightforward to
  derive from the definitions that
  \[
  \D(t^{-}) = \Du(t^{-}) \land \Ddm{t}(t^{-}) \land \dia\Dd(t),
  \]
  and
  \[
  \D(t) = \diap(\Du(t^{-}) \land \Ddm{t}(t^{-})) \land \Dd(t).
  \]
  Hence, \eqref{eq:4c} follows from the conjugacy of $\dia$ and
  $\diap$: simply take $a = \Du(t^{-}) \land \Ddm{t}(t^{-})$ and $b
  = \Dd(t)$ in \eqref{eq:cj}.  
\end{proof}
Call a finite network $N = \struc{T,L}$ \emph{globally coherent} if
$\D(t)>\bot$ for all $t\in T$.  We can now prove our \emph{repair
  lemma}.  We say that $N'$ \emph{extends} $N$, notation: $N \leq N'$,
if $T \subseteq T'$ and $L(t) \subseteq L'(t)$ for every $t \in T$.

\begin{lemma}[Repair Lemma]
\label{l:4:4}
Let $N = \struc{T,L}$ be a globally coherent $A$-network.
Then for any defect $d$ of $N$ there is a globally coherent extension $N^{d}$
of $N$ which lacks the defect $d$.
\end{lemma}

\begin{proof}
  We will take action depending on the type of the defect $d$.  In
  each case we will make heavily use of the global extension $\D$ of
  $L$.
  \begin{enumerate}
  \item If $d = (t,a,\cp)$ is a defect of the first kind, then we
    define $N^{d} := \struc{T,L^{d}}$, where $L^{d}(s) := L(s)$ for $s
    \neq t$, while we put
    \[
    L^{d}(t) := \left\{ \begin{array}{ll}
        L(t) \cup \{ a     \} & \mbox{ if } \D^{N}(t) \land a > \bot,
        \\ L(t) \cup \{ \cp a \} & \mbox{ if } \D^{N}(t) \land \cp a > \bot.
      \end{array}\right.
    \]
    Then clearly the triple $(t,a,\cp)$ is no longer a defect, and so
    all that is left to show is the global coherence of $N^{d}$.  But
    since $\D^{N}(t) > \bot$ by assumption, we will have either
    $\D^{N}(t) \land a > \bot$ or $\D^{N}(t) \land \cp a> \bot$.  It
    is easy to check that in either case, we have $\D^{N^{d}}(t) =
    \D^{N}(t) \land x$ with $x \in \{ a,\cp a\}$, and from this
    coherence follows easily.
 
  \item Now suppose that $d = (t,a,\dia)$ is a type 2 defect.  Let $k$
    be the least number such that $tk \not\in T$, and define $N^{d} :=
    \struc{T^{d}, L^{d}}$, where $T^{d} = T \cup \{ tk \}$, and
    $L^{d}$ is given by putting $L^{d}(s) := L(s)$ for $s \neq t$,
    while $L^{d}(tk) := \{ a \}$.  In this case it is easy to prove
    that $\D^{N^{d}}(t) = \D^{N}(t)$, so $N^{d}$ is certainly globally
    coherent.  It is likewise simple to see that $(t,a,\dia)$ is no
    longer a defect of $N$.

  \item Finally, suppose that $d = (t,\vec{a},\ff)$ is a defect of the
    third kind.  By global coherency we have that $\D^{N}(t) > \bot$.
    Suppose for contradiction that $\D^{N}(t) \land
    (\ga_{\vec{a}}^{A})^{n}(\bot) = \bot$ for all numbers $n$.  Then
    for all $n$ we have $(\ga_{\vec{a}}^{A})^{n}(\bot) \leq \cp
    \D^{N}(t)$, and so by constructiveness of \ff\ on $A$ it follows
    that $\ff^{A}\vec{a} \leq \cp\D^{N}(t)$.  But this contradicts the
    fact that $N$ is coherent.

    It follows that $\D^{N}(t) \land (\ga_{\vec{a}}^{A})^{n}(\bot) >
    \bot$ for some natural number $n$.  Now proceed as in the first
    case, defining $L^{d}(t) := L(t) \cup \{
    (\ga_{\vec{a}}^{A})^{n}(\bot) \}$.
  \end{enumerate}
\end{proof}

\begin{lemma}
\label{l:4:5}
Every globally coherent $A$-network can be extended to a perfect
network.
\end{lemma}

\begin{proof}
We will define a sequence of networks $N = N_{0} \leq N_{1} \leq
N_{2} \leq \ldots\ $ such that for each $i\in\om$ and each defect
$d$ of $N_{i}$ there is a $j>i$ such that $d$ is not a defect of
$N_{j}$.

For the details of this construction, define
\[
D := \om^{*} \times A \times \{\cp,\dia\} \,\cup\, 
   \om^{*} \times A^{n} \times \{\ff\}.
\]
Informally we shall say that $D$ is the set of \emph{potential defects}.
Clearly, since $D$ is countable, we may assume the
  existence of an enumeration $(d_{n})_{n<\om}$ such that every
  element of $D$ occurs infinitely often.

  Now we set
  \[\begin{array}{lll}
    N_{0} &:=& N
    \\ N_{i+1} &:=& \left\{\begin{array}{ll}
        N_{i}^{d_{i}} & \mbox{if $d_{i}$ is actually a defect of $N_{i}$}\,,
        \\ N_{i}      & \mbox{otherwise}\,.
      \end{array}\right.
  \end{array}\]
  Finally, define $N' := \struc{T',L'}$, with $T':= \bigcup_{i<\om}
  T_{i}$ and, for each $t \in T'$, $L'(t) := \bigcup_{i<\om} L_{i}(t)$.
  It is then straightforward to verify that $N'$ is a perfect extension of $N$.
For instance, suppose for contradiction that $N'$ would have some
  defect $d$.  It readily follows from the definitions that there must
  be some approximation $N_{k}$ in the sequence for which $d$ is also
  a defect.  But then the next time $i$ such that $d = d_{i}$, this
  defect will be repaired.  As a consequence, $d$ is not a defect of
  $N_{d_{i}+1}$, and so it cannot be a defect of $N'$ either.  This
  provides the desired contradiction.
\end{proof}

\begin{proof}[Proof of Lemma~\ref{l:4:1}]
  Consider an arbitrary nonzero element $a \in A$, and let $N_{a}$ be
  the network $\struc{\{\eps\},L_{a}}$, $L_{a}$ given by $L_{a}(\eps)
  := \{ a\}$.  It is obvious that $N_{a}$ is globally coherent, so
  Lemma~\ref{l:4:1} follows by a direct application of the
  Lemmas~\ref{l:4:5} and~\ref{l:4:2}.
\end{proof}

\begin{proof}[Proof of Theorem~\ref{\reftotheo}]
  Let $S$ be the disjoint union of the family $\set{ S_{a} \mid
    \bot\neq a \in A }$, where for each nonzero $a \in A$, $S_{a}$ is
  given by Lemma~\ref{l:4:1}.  It is straightforward to verify that
  $A$ can be embedded into the product $\prod_{a\neq\bot}S_{a}^{\ff}$,
  and that this latter product is isomorphic to $S^{\ff}$, the complex
  $\ff$-algebra of $S$.
\end{proof}


\bibliographystyle{elsart-num-sort}
\bibliography{bibliography}

\def\ocirc#1{\ifmmode\setbox0=\hbox{$#1$}\dimen0=\ht0 \advance\dimen0
  by1pt\rlap{\hbox to\wd0{\hss\raise\dimen0
  \hbox{\hskip.2em$\scriptscriptstyle\circ$}\hss}}#1\else {\accent"17 #1}\fi}
\begin{thebibliography}{10}
\expandafter\ifx\csname url\endcsname\relax
  \def\url#1{\texttt{#1}}\fi
\expandafter\ifx\csname urlprefix\endcsname\relax\def\urlprefix{URL }\fi

\bibitem{lics2005}
Proceedings of the Twentieth IEEE Symposium on Logic in Computer Science (LICS
  2005), IEEE Computer Society Press, 2005 (2005).

\bibitem{arnoldniwinski}
A.~Arnold, D.~Niwi{\'n}ski, Rudiments of {$\mu$}-calculus, No. 146 in Studies
  in Logic and the Foundations of Mathematics, North-Holland Publishing Co.,
  Amsterdam, 2001.

\bibitem{barw:vici96}
J.~Barwise, L.~Moss, Vicious Circles, vol.~60 of CSLI Lecture Notes, CSLI
  Publications, 1996.

\bibitem{bilk:proo08}
M.~B\'{\i}lkov\'a, A.~Palmigiano, Y.~Venema, Proof systems for the coalgebraic
  cover modality, in: C.~Areces, R.~Goldblatt (eds.), Advances in Modal Logic,
  Volume 7, College Publications, 2008.

\bibitem{blac:moda01}
P.~Blackburn, M.~{de R}ijke, Y.~Venema, Modal Logic, No.~53 in Cambridge Tracts
  in Theoretical Computer Science, Cambridge University Press, 2001.

\bibitem{bloomesik}
S.~L. Bloom, Z.~{\'E}sik, Iteration theories, Springer-Verlag, Berlin, 1993.

\bibitem{boffa}
M.~Boffa, Une condition impliquant toutes les identit\'es rationnelles, RAIRO
  Informatique Th\'eorique et Applications 29~(6) (1995) 515--518.

\bibitem{burg:axio82}
J.~Burgess, Axioms for tense logic {I}: `since' and `until', Notre Dame Journal
  of Formal Logic 23 (1982) 375--383.

\bibitem{conway}
J.~Conway, Regular Algebra and Finite Machines, Chapman and Hall, 1971.

\bibitem{emer:deci85}
E.~Emerson, J.~Halpern, Decision procedures and expressiveness in the temporal
  logic of branching time, Journal of Computer and System Sciences 30 (1985)
  1--24.

\bibitem{emerson}
E.~A. Emerson, Temporal and modal logic, in: Handbook of theoretical computer
  science, Vol.\ B, Elsevier, Amsterdam, 1990, pp. 995--1072.

\bibitem{faginhalpernetal}
R.~Fagin, J.~Y. Halpern, Y.~Moses, M.~Y. Vardi, Reasoning about knowledge, MIT
  Press, Cambridge, MA, 1995.

\bibitem{fine:norm75}
K.~Fine, Normal forms in modal logic, Notre Dame Journal of Formal Logic 16
  (1975) 229--234.

\bibitem{kozen:PDL}
D.~Harel, D.~Kozen, J.~Tiuryn, Dynamic Logic, MIT Press, Cambridge, MA, 2000.

\bibitem{hodk:rela00}
R.~Hirsch, I.~Hodkinson, Relation Algebras by Games, No. 147 in Studies in
  Logic and the Foundations of Mathematics, North-Holland Publishing Co.,
  Amsterdam, 2002.

\bibitem{hodg:mode93}
W.~Hodges, Model Theory, Cambridge University Press, 1993.

\bibitem{jani:auto97}
D.~Janin, Automata, tableaus and a reduction theorem for fixpoint calculi in
  arbitrary complete lattices, in: Proceedings of the Twelfth Annual IEEE
  Symposium on Logic in Computer Science (LICS 1997), IEEE Computer Society
  Press, 1997.

\bibitem{jani:auto95}
D.~Janin, I.~Walukiewicz, Automata for the modal $\mu$-calculus and related
  results, in: Proc.~MFCS'95, Springer, Berlin, 1995, lNCS 969.

\bibitem{kozen}
D.~Kozen, Results on the propositional {$\mu $}-calculus, Theoretical Computer
  Science 27~(3) (1983) 333--354.

\bibitem{kozen:regexpr}
D.~Kozen, A completeness theorem for {K}leene algebras and the algebra of
  regular events, Inform. and Comput. 110~(2) (1994) 366--390.

\bibitem{koze:elem81}
D.~Kozen, R.~Parikh, An elementary proof of the completeness of {PDL},
  Theoretical Computer Science 14 (1981) 113--118.

\bibitem{krob}
D.~Krob, Complete systems of {B}-rational identities, Theoret. Comput. Sci.
  89~(2) (1991) 207--343.

\bibitem{kupk:comp08}
C.~Kupke, A.~Kurz, Y.~Venema, A complete coalgebraic logic, in: C.~Areces,
  R.~Goldblatt (eds.), Advances in Modal Logic, Volume 7, College Publications,
  2008.

\bibitem{kupk:clos05}
C.~Kupke, Y.~Venema, Closure properties of coalgebra automata, in: LICS 2005
  \cite{lics2005}, pp. 199--208.

\bibitem{kupk:coal08}
C.~Kupke, Y.~Venema, Coalgebraic automata theory: basic results, Logical
  Methods in Computer Science.

\bibitem{lang:focu01}
M.~Lange, C.~Stirling, Focus games for satisfiability and completeness of
  temporal logic, in: Proceedings of the Sixteenth Annual IEEE Symposium on
  Logic in Computer Science (LICS 1997), IEEE Computer Society Press, 2001.

\bibitem{lynd:repr50}
R.~Lyndon, The representation of relation algebras, Annals of Mathematics 51
  (1950) 707--729.

\bibitem{madd:some82}
R.~Maddux, Some varieties containing relation algebras, Transaction of the
  American Mathematical Society 272 (1982) 501--526.

\bibitem{meye:epis95}
J.-J. Meyer, {W. van der H}oek, Epistemic Logic for AI and Computer Science,
  No.~41 in Cambridge Tracts in Theoretical Computer Science, Cambridge
  University Press, 1995.

\bibitem{moss:coal99}
L.~Moss, Coalgebraic logic, Annals of Pure and Applied Logic 96 (1999)
  277--317, (Erratum published \textit{Ann.P.Appl.Log.} 99:241--259, 1999).

\bibitem{palm:nabl07}
A.~Palmigiano, Y.~Venema, Nabla algebras and {C}hu spaces, in: Algebra and
  Coalgebra in Computer Science (CALCO 2007), Springer-Verlag, Berlin, 2007,
  lNCS 4624.

\bibitem{rutt:univ00}
J.~Rutten, Universal coalgebra: {A} theory of systems, Theoretical Computer
  Science 249 (2000) 3--80.

\bibitem{santocanale:eqfixedpoints}
L.~Santocanale, On the equational definition of the least prefixed point,
  Theoretical Computer Science 295~(1-3) (2003) 341--370.

\bibitem{santocanale:LICS05}
L.~Santocanale, Completions of $\mu$-algebras, in: LICS 2005 \cite{lics2005},
  pp. 219--228.

\bibitem{sant:comp08}
L.~Santocanale, Completions of $\mu$-algebras, Annals of Pure and Applied Logic
  154 (2008) 27--50.

\bibitem{sege:comp82}
K.~Segerberg, A completeness theorem in the modal logic of programs, in:
  T.~Traczyk (ed.), Universal Algebra and Applications, vol.~9 of Banach Centre
  Publications, PWN--Polish Scientific Publishers, 1982, pp. 31--46.

\bibitem{tarski}
A.~Tarski, A lattice-theoretical fixpoint theorem and its applications, Pacific
  Journal of Mathematics 5 (1955) 285--309.

\bibitem{vene:alge06}
Y.~Venema, Algebras and coalgebras, in: P.~Blackburn, J.~van Benthem, F.~Wolter
  (eds.), Handbook of Modal Logic, Elsevier, 2006, pp. 331--426.

\bibitem{vene:auto06}
Y.~Venema, Automata and fixed point logic: a coalgebraic perspective,
  Information and Computation 204 (2006) 637--678.

\bibitem{walukiewicz}
I.~Walukiewicz, Completeness of {K}ozen's axiomatisation of the propositional
  {$\mu$}-calculus, Inform. and Comput. 157~(1-2) (2000) 142--182, (An earlier
  version of the proof appeared in the Proceedings of LICS 1995.).

\end{thebibliography}

\end{document}